%% file: main.tex
\documentclass[preprint,3p,singlecolumn]{elsarticle}

\usepackage{lineno,hyperref}
\modulolinenumbers[5]
\usepackage{xspace} 
\usepackage{siunitx} 
\usepackage{wasysym}
\usepackage{color}

\journal{Nuclear Instruments and Methods in Physics Research A}

\newcommand{\fpa}{fixed-NMR-probe array\xspace}

\newcommand{\trolley}{\textsl{magnetic field trolley}\xspace}
\newcommand{\msv}{muon storage volume\xspace}
\newcommand{\sh}{\textsl{sample holder}\xspace}
\newcommand{\bp}{\textsl{base piece}\xspace}
\newcommand{\ic}{\textsl{inner crimp ring}\xspace}
\newcommand{\oc}{\textsl{outer crimp ring}\xspace}
\newcommand{\icond}{\textsl{inner conductor}\xspace}
\newcommand{\tol}[4]{\ensuremath{\num{#1}^{+\num{#2}}_{-\num{#3}}}\,\si{#4}}
\newcommand{\omegaL}{\ensuremath{\omega_\mathrm{L}}}
\newcommand{\nuL}{\ensuremath{\nu_\mathrm{L}}}

\begin{document}
\input{0_title_author_abstract_keyword}


\input{1_intro}
\input{2_NMR}

\input{3_fixedprobes}

\input{4_electronics}

\input{5_performance}

\input{6_conclusion}

\input{7_acknowledgement}

\input{8_Appendices}

\newpage
\bibliography{main.bib}

\end{document}

%% file: 0_title_author_abstract_keyword.tex
\begin{frontmatter}
\title{The fixed probe storage ring magnetometer for the Muon g-2 experiment at Fermi National Accelerator Laboratory}

\author[uw]{E.\,~Swanson\corref{ESnote}}
\cortext[ksknote]{Corresponding author: Tel.: +1-206-543-4080.}
\ead{eriks@uw.edu}
\author[uw,jgu]{M.\,~Fertl} 
\author[uw]{A.\,~Garcia}
\author[uw]{C.\,~Helling}
\author[uw]{R.\,~Ortez}
\author[uw]{R.\,~Osofsky}
\author[uw]{D.\,A.~Peterson}
\author[jgu]{R.~Reimann}
\author[uw]{M.\,W.~Smith}
\author[uw]{T.\,D.~Van\,Wechel}

\affiliation[uw]{organization={University of Washington}, addressline={Box 351560}, postcode={Seattle, WA 98195},country = {USA}}
\affiliation[jgu]{organization= {Johannes Gutenberg University Mainz}, postcode ={55128 Mainz}, country={Germany}}

\begin{abstract}
The goal of the FNAL E989 experiment is to measure the muon magnetic anomaly to unprecedented accuracy and precision at the Fermi National Accelerator Laboratory. 
To meet this goal, the time and space averaged magnetic environment in the muon storage volume must be known to better than \SI{70}{ppb}. 
A new pulsed proton nuclear magnetic resonance (NMR) magnetometer was designed and built at the University of Washington, Seattle to track the temporal stability of the \SI{1.45}{\tesla} magnetic field in the muon storage ring at this precision. 
It consists of an array of 378 petroleum jelly based NMR probes that are embedded in the walls of muon storage ring vacuum chambers and custom electronics built with readily available modular radio frequency (RF) components. 
We give NMR probe construction details and describe the functions of the custom electronic subsystems. 
The excellent performance metrics of the magnetometer are discussed, where after 8 years of operation the median single shot resolution of the array of probes remains at 650 ppb.

\end{abstract}

\begin{keyword}
pulsed nuclear magnetic resonance, magnetic field, muon, anomalous magnetic moment
\end{keyword}

\end{frontmatter} 

%% file: 1_intro.tex
\section{Introduction}
\label{sec:intro}
\begin{figure}
    \centering
    \includegraphics[width=0.9\linewidth]{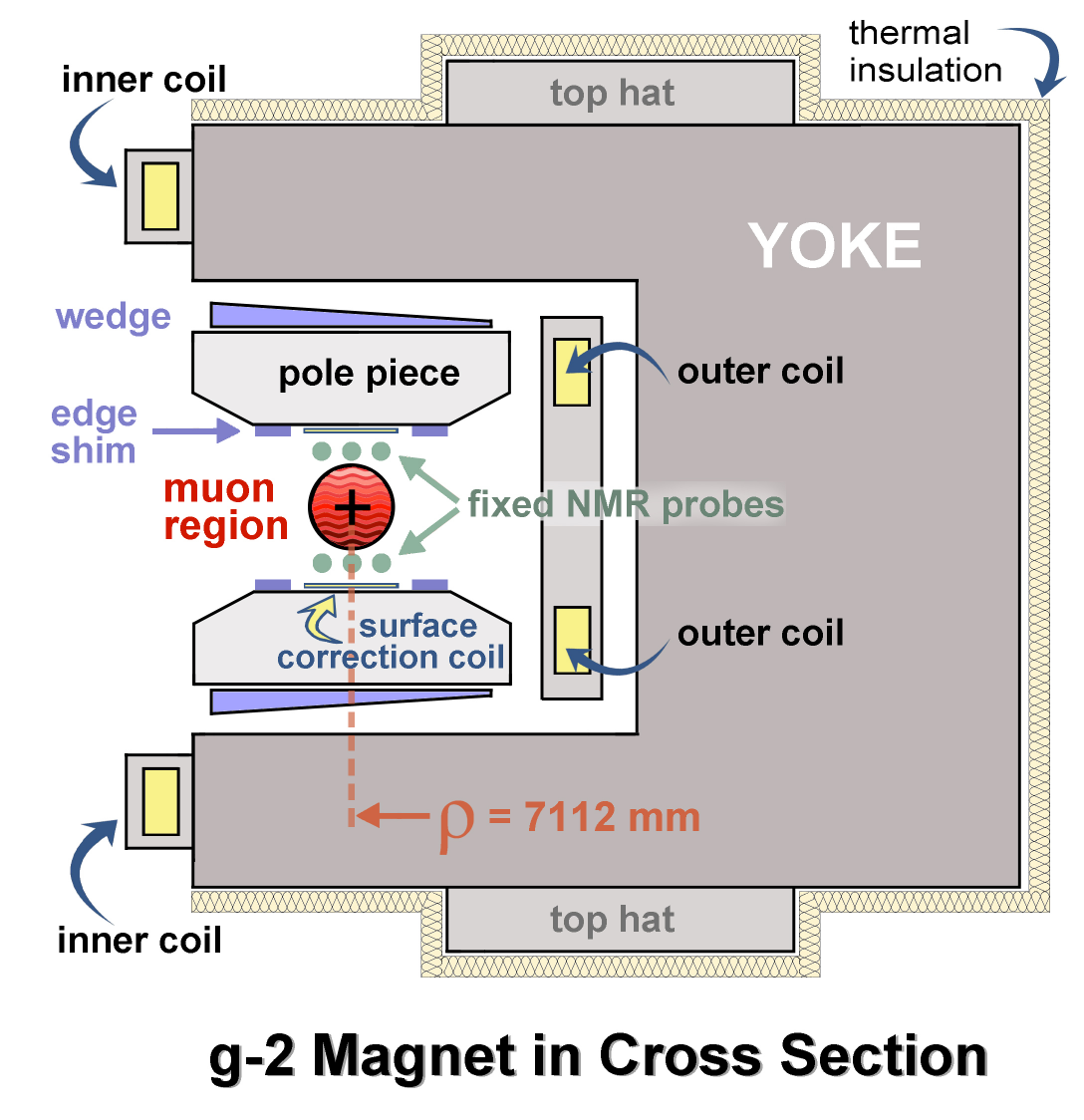}
    \caption{Cross-section view of the super-ferric muon g-2 storage ring magnet. The open part of the magnet's yoke faces the center of the ring. Between the upper and lower pole pieces is the muon storage region with magnetometer NMR probes located above and below. Pole pieces are individually shimmed to produce a homogeneous azimuthal magnetic field. }
    \label{fig:Fermiring}
\end{figure}

The measurement of the muon magnetic anomaly, $a_\mathrm{\mu}$ tests the predictions of the Standard Model of Particle Physics (SM)  and relies on a high accuracy determination of the magnetic field in the muon storage ring.
The BNL E821 experiment at Brookhaven National Laboratory, Upton, New York, USA,  measured $a_\mathrm{\mu}$ 
between 1997 and 2001 \cite{PhysRevD2003} with an accuracy of \SI{550}{ppb}.
E821 employed a super-ferric storage ring magnet, shown in cross-section in figure~\ref{fig:Fermiring}.
The super conducting magnet operates in non-persistant mode at a field of 1.45 T in the toroidal muon storage volume (major radius R = 7.112 m and minor radius r= 4.5 cm).
The pulsed proton NMR systems developed for and employed in the BNL E821 experiment consist of an in-vacuum  \trolley and a system of distributed NMR probes at fixed locations above and below the muon storage region (shown in figure~\ref{fig:Fermiring}) for continuously monitoring the field. 
The trolley is deployed throughout the \msv and maps the magnetic field when muons are not being stored in the ring. 
Magnetometer field measurements are given in terms of proton spin precession frequencies. 
These come from directly averaging zero crossing time intervals of free induction decay (FID) waveforms in hardware. 
E821 systems are further described in reference \cite{PRIGL1996}.
A separate precision NMR magnetometer, based on well known properties of protons in water \cite{Phillips_1977}, provided the absolute field calibration for the measurement. 
In 2013, the super-ferric muon storage ring was  transported to Fermi National Accelerator Laboratory, Batavia, Illinois, USA to serve a second life as the heart of the new FNAL E989 experiment. 
The new experiment aims to achieve a 4-fold reduction in uncertainty (\SI{140}{ppb}) on $a_\mathrm{\mu}$. 
To achieve this goal, the measurement precision and accuracy of the \SI{1.45}{\tesla} magnetic field  will need a twofold improvement over the E821 measurement \cite{TDR2015}.
The FNAL E989 \textsl{fixed-probe magnetometer} is conceptually similar to the one deployed in E821, but with redesigned NMR probes and electronics. 
In contrast to E821, the magnetometer provides actual digitized FID waveforms for extracting precession frequencies by subsequent analysis to better understand the magnetic environments of the probes \cite{HONG2021107020}. 
Here we present design details of the new fixed probe magnetometer, the redesigned NMR probes and associated  pulsed radio-frequency electronics. 
The in-vacuum trolley was refurbished and significantly upgrade, as described in reference \cite{Corrodi_2020}. 
The NMR probes (described in section \ref{sec:NMRProbeHeads}) were also used for the trolley system. 
Lastly, reference \cite{Flay_2021} describes the design and development of the new water-based probe used for absolute magnetic field calibration in E989. 
The BNL absolute water-based probe has been cross-calibrated by a hyper-polarized $^3$He probe~\cite{farooq_absolute_2020} and similar measurements for the E989 is in preparation. 

%% file: 2_NMR.tex
\section{Nuclear magnetic resonance}
\label{sec:NMR}
The accurate determination of the storage ring magnetic field  relies on pulsed proton NMR. The technique used here is best described in storage ring magnet coordinates.
Observing the ring in cross-section in figure~\ref{fig:Fermiring},
the main field component $\vec{B}_{\mathrm{0}}$ lies in the vertical $y$-direction (up in the figure) of a locally Cartesian coordinate system. 
The positive $x$ direction is pointing away from the center of the ring (right in the figure) and $z$ points out of the page in the direction of travel of the muon beam.
For protons exposed to a magnetic flux density $B_0$, the Larmor frequency is $\nuL = B_0 \frac{\gamma_p^{\prime}}{2\pi} $, where $\gamma_p^{\prime}$ is the gyro-magnetic ratio of the shielded protons in the sample material. 
For a single free proton $\gamma_p/2\pi= \SI{42.577 478 518(18)}{MHz\per\tesla}$ \cite{CODATA2018}.
Here we employ the classical vector description of NMR \cite{Keeler} in which the macroscopic equilibrium magnetization $\vec{M}$ of the proton sample is given by  
\begin{equation}
    \vec{M} = P_0 N \mu_p \hat{{B}}_0,
\end{equation}
where $N$ is the number density of magnetic moments, $P_0$ is the polarization of the material, $\mu_p$ the magnetic moment of the individual proton in the sample material and $\hat{B}_0$ the unit vector in the direction of the magnetic field.
The orientation of $\vec{M}$ can be manipulated with oscillating magnetic fields of frequency $\omega$. 
The effect of oscillating magnetic fields can be conveniently described in a coordinate system co-rotating with frequency $\omega$ around $\hat{B}$. 
For a magnetic radio frequency (RF) field $\vec{B}_1$ oriented perpendicular to $\vec{B}_0$ the  magnetic field $\vec{B}_\mathrm{cr}$ in the co-rotating coordinate system is given by
\begin{equation}
    \vec{B}_\mathrm{cr} = \vec{B}_1 + \vec{B}_0\left(1-\frac{\omega}{\omegaL}\right).
\end{equation} 
On resonance, $\omega = \omegaL$ the contribution of the static external magnetic field $\vec{B}_0$ vanishes.
Then the direction of the effective magnetic field $\vec{B}_\mathrm{cr}$ is given by $\hat{B}_1$.
In the co-rotating frame, the magnetization vector will in general be rotating around the direction of $\vec{B}_\mathrm{cr}$.
If the RF pulse is applied on resonance and for the time  $t_\frac{\pi}{2} = \frac{\pi}{2\gamma_p B_1}$ the magnetic moments will be tilted by $\pi/2$ into the x-y plane of the co-rotating frame. A pulse of this duration is referred to as the $\pi /2$ pulse.
In the laboratory frame, the magnetization will thus precess with the Larmor frequency $\omegaL/2\pi$.
The general time evolution of the magnetization vector can be described by the Bloch equations \cite{Bloch1946}
\begin{equation}
    \frac{M_{x,z}}{d t} = \gamma_p \left( \vec{M} \times \vec{B}\right)_{x,z}-\frac{M_{x,z}}{T_2} 
\end{equation}
and
\begin{equation}
    \frac{M_{y}}{d t} = \gamma_p \left(\vec{M} \times \vec{B}_0\right)_y - \frac{M_y-M_0}{T_1}
\end{equation}
with the longitudinal relaxation time $T_1$ and the transverse relaxation time $T_2$. $T_1$ governs the maximum measurement repetition rate.
%
%
The proton sample is surrounded by a coil which forms the inductor of a series resonant LC circuit. 
The coil acts as the excitation coil when delivering the RF pulse and as a pick-up coil in which the excited precession of $\vec{M}$ induces an electromotive force according to Faraday's law.
The resulting waveform is the free induction decay or FID.
%

%% file: 3_fixedprobes.tex
\section{NMR probes for the \fpa and trolley}
\label{sec:NMRProbeHeads}
Detailed visual inspections and electrical tests of the NMR probes employed by the BNL E821 experiment revealed that most of these probes had failed from mechanical and/or electrical reasons by the time they had been removed from the E821 setup, or potentially even earlier.
E821 probes used water for their proton rich material, doped with $\mathrm{CuSO_4}$ to shorten the $T_1$ relaxation time. 
Leaks and evaporation over time affected the FID signals.
To meet performance goals, new substantially upgraded NMR probes for both the \fpa and the \trolley were developed and built to address the identified failure modes. 
The NMR probes described here use petroleum jelly for their proton rich material. 
Petroleum jelly has a relaxation time $T_1$ of \SI{30.5}{\milli\second} which is comparable to that of doped water and  does not evaporate. 
NMR studies done at the University of Washington at high fields resolved two separate proton frequencies in petroleum jelly, but that difference is small enough that they are not resolved at the \SI{1.45}{T} ring field. 
The measured temperature dependence was a few ppb/K. 
The nominal storage ring field was determined by muon beam dynamics, and probe tuning was limited in range to a few percent of its nominal value by design.

\subsection{Mechanical and electrical design considerations}
An important design consideration of NMR probes used in both the \fpa and the \trolley was their long-term stability and reliability
over the multi-year physics data taking period of FNAL E989.
Accessing individual probes in the \fpa would only be possible by removing a complete vacuum chamber section from the experiment.
To access individual \trolley probes, the \msv first would need to be brought to atmospheric pressure and the \trolley removed and unsealed. Any changes in probe positions from this process invalidates their calibrations.
Both procedures would be prohibitively time-consuming.
The implemented design upgrades include
\begin{itemize}
    \item a more stable mechanical layout that allows frequency tuning of the resonance circuit without inducing mechanical stress on the probe assembly.
    \item electrical crimp connections on the cable and on the parallel inductor, which reduces contact resistance and increases mechanical and electrical long-term stability.
    \item ultrasonic soldering of aluminum to copper connections.
    \item the exclusive use of non-corrosive petroleum jelly as the proton sample, which avoids proton sample leakage (resulting in signal loss) and corrosion of probe construction materials. 
\end{itemize}
The exterior dimensions of the probes were limited by the milled-in slots of the E821 vacuum chambers, which were re-used in E989.
The choice of construction material was strictly limited to non-ferromagnetic materials. 
Paramagnetic and diamagnetic materials have magnetization that scale with the applied field.
\subsection{Mechanical design details}
\label{subsec:MechanicalDesignDetails}
\begin{figure*}[htbp]
\centering
\includegraphics[width=\linewidth]{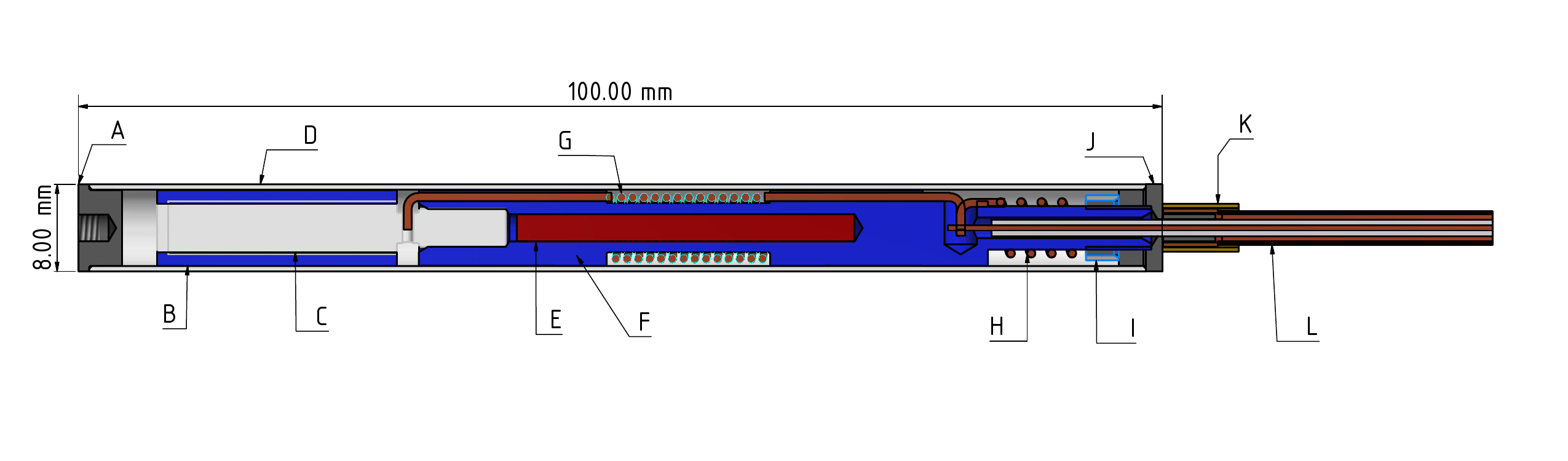}
\caption{Cut-away view of the mechanical design of the NMR probe. The mechanical backbone of the probe is the \sh (F), made from PTFE, which provides mechanical stability to the serial (G) and parallel (H) inductor coil. The \sh accommodates the petroleum jelly sample (E) which is sealed off by the \icond (C) of the tune-able cylinder capacitor. Its capacitance is determined by the dielectric fill factor between the inner conductor and the aluminum sleeve (D) acting as the outer conductor.  The position of the PTFE dielectric (B) is used to change the fill factor of the capacitor.
The parallel inductor coil is mechanically and electrically connected to the \bp (J) with the \ic (I). The \oc (K) is used to crimp the outer conductor of the double shielded RF cable (L) to the \bp. The NMR probe is mechanically closed with an end cap (A).}
\label{fig:FixedProbeBody}
\end{figure*}

Here we present details of the mechanical design. 
A detailed description of the assembly process of an individual NMR probe is given in appendix\,\ref{subsec:ProbeAssemblyProcess}.
A cut away view of the mechanical design of the NMR probe body is shown in figure \ref{fig:FixedProbeBody}. 
Each NMR probe has a cylindrical shape of \SI{8.00}{\milli\meter} outer diameter (OD) and a total length of \SI{100.0}{\milli\meter}.
The inner mechanical backbone of the NMR probe is provided by the  \sh. 
Made from polytetrafluoroethylene (PTFE), the \sh accommodates a coaxial cylindrical cavity of \SI{2.5}{\milli\meter} diameter and \SI{31.7}{\milli\meter} length filled with petroleum jelly as the active proton sample. 
A 32-turn coil (Essex, GP/MR-200, AWG 30 magnet wire, copper) is wound around the center of the bubble-free petroleum jelly sample into cut grooves structured with a M4.5x0.5 external thread of \SI{15.00}{\milli\meter} length (thread outer diameter \SI{4.60}{\milli\meter}). 
This coil is the inductor of the series LC circuit.
The connecting ends of the inductor run embedded under the surface of the \sh in a straight groove (depth \tol{0.99}{0.03}{0.00}{\milli\meter}, width \SI{0.28}{\milli\meter}) to preserve the cylindrical surface and to avoid electrical shorts to the aluminum sleeve. 
At the top of the \sh, a double-threaded aluminum (Al6061) cylinder, screwed into an M3 thread, seals the petroleum jelly sample cavity. 
One end of the coil is soldered into a wire eye formed from a bare cooper wire. 
The bare copper wire is soldered into a through hole in the aluminum conductor using an ultrasonic soldering technique \cite{MBR} for low resistance electrical contact and longevity.
The lower end of the \sh a coaxial recess (OD \SI{4.00}{\milli\meter}, length \SI{9.00}{\milli\meter}) accommodates a second parallel inductor coil used to impedance-match the driven end of the LC circuit to the \SI{50}{\ohm} impedance of the double-shielded non-magnetic RF cable (Huber \& Suhner GS 02232 D). 
The \sh screws into a M4x0.5 internal thread (depth \SI{6.00}{\milli\meter}) in the \bp around the dielectric and inner conductor of the RF cable, which enters the NMR probe through coaxial holes in the \bp (diameter \SI{1.6}{\milli\meter}) and the \sh (diameter \SI{1.9}{\milli\meter}). 
The \bp provides the electrical connections for the RF cable outer conductor and the parallel inductor coil. 
For good electrical and mechanical stability, both connections are crimped. 
The bare end of parallel inductor coil wire (Essex, GP/MR-200, AWG 28, bare diameter \SI{0.32}{\milli\meter}) is placed in a partially cut open off-axis hole (center radial position \SI{2.45}{\milli\meter}, diameter \SI{0.4}{\milli\meter}) in a cylindrical step (diameter \SI{5.00}{\milli\meter}) in the \bp. 
The wire is secured in the hole in the radial direction while the wire's surface rises \SI{0.11}{\milli\meter} above the cylindrical surface of the \bp and is crimped in place with an \ic. 
This \ic (inner diameter (ID) \SI{5.00}{\milli\meter}, OD \SI{6.00}{\milli\meter}, length \SI{3.00}{\milli\meter}, Al1100) is slightly undersized compared to the maximum dimension across the assembly dimension at the wire location of \SI{5.15}{\milli\meter}. 
During the installation, the \ic is pressed over the wire/base piece connection which pinches the wire permanently in place. 
The applied force also breaks up the aluminum oxide layer, which reduces the local contact resistance.
The outer conductor of the double-shielded cable is mechanically and electrically connected by crimping the  \oc (ID \SI{3.60}{\milli\meter}, OD \SI{4.40}{\milli\meter}, length \SI{7.0}{\milli\meter}, Al1100) onto the RF cable feed-through section (OD \SI{2.70}{\milli\meter}) of the \bp. 
The assembly of \sh, \bp and RF cable is inserted into a precision sleeve (OD \SI{8.00\pm 0.10}{\milli\meter}, ID \tol{7.000}{0.005}{0.000}{\milli\meter}, length \SI{97.00}{\milli\meter}) made from \SI{99.5}{\percent} aluminum bought from \url{alu-verkauf.de}. 
The tight fit between the aluminum sleeve and the \bp (OD of mating surface \tol{7.000}{0.000}{0.010}{\milli\meter}) provides mechanical stability and good electrical contact. 
The latter is significant because the aluminum sleeve  influences the inductance of the series inductor coil (due to induced mirror surface currents) and is also the outer conductor of the tuneable cylindrical capacitor of the LC circuit. 
Its capacitance is adjusted by changing its dielectric filling factor. 
The position of a thin PTFE cylinder with internal thread (M5x0.5) can be adjusted along the length of the \icond with an external thread (M5x0.5). 
For good position control, the \icond has a rectangular groove in its end face with allows locking its angular position in place while rotating the PTFE dielectric with a custom screwdriver. 
The hollow screwdriver clears the \icond and engages with two teeth in two grooves in the front face of the PTFE dielectric. 
The development of the special screwdriver was key to performing the PTFE position adjustment after the NMR probe was fully assembled, without risk of breaking the solder joint between the \ic and the parallel inductor coil wire.
The NMR probe is physically and electrically closed by an aluminum cap. 
Due to the proximity of the cap to the \icond there is a small extra contribution to the capacitance of the capacitor.
A threaded hole (M3x0.5) in the outer face allows for easy removal of the cap for the frequency tuning procedure.
To that end, a plastic screw is inserted in the thread. The plastic screw is used to co-axially pull out the end cap to avoid deformation of the soft precision aluminum sleeve. 

%% file: 4_electronics.tex
\section{The pulsed NMR electronics system}
\label{sec:NMRElectronics}

\begin{figure}
    \includegraphics[width=1.0\linewidth]{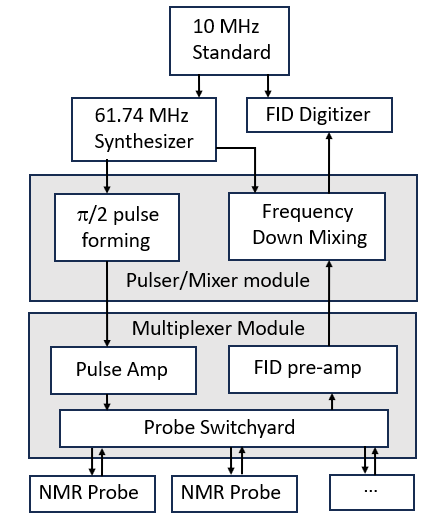}
   \caption{Block diagram showing the NMR magnetometer. At the top are the GPS disciplined  reference, the 61.74 MHz frequency synthesizer and waveform digitizer. The two shaded blocks are  pulser/mixer and multiplexer modules. The pulser/mixer generates the $\pi$/2 excitation pulse and down converts the returning FID for digitization. An external trigger simultaneously initiates the $\pi$/2 pulse and starts the waveform digitizer. In the multiplexer module, the selected probe is first connected to the pulse amplifier to excite the resonance and then to the preamplifier which sends the FID back to the mixer.}
   \label{fig:ElectronicsScheme}
\end{figure}

The NMR electronic systems generate and deliver RF excitation pulses to the 378 NMR probes and receive and process FID signals before they are digitized for further analysis.
Figure \ref{fig:ElectronicsScheme} shows a block diagram of the individual components of the NMR magnetometer. 
The design employs two custom electronic modules:  {\it pulser-mixers} which are 1-U-wide NIM modules located in the low-magnetic field region at the ring center and  {\it multiplexers}, stand-alone units on top of the storage ring magnet that connect to individual probes. 
These are shown within the shaded regions of  figure~\ref{fig:ElectronicsScheme}.
Modular RF components, readily available from MiniCircuits (MC), are used as building blocks for magnetometer circuits.  
Individual part numbers given in the text are from the Mini-Circuits catalog~\cite{Minicircuit:Catalog}.  
Digital logic elements use standard transistor-transistor logic (TTL) components with part numbers given in the text. 
The pulsers (the pulse shape forming box in the figure) form the timed burst of \SI{61.74}{MHz} RF ($\pi$/2 pulse) from a clock synthesizer and send them to multiplexers where they are further amplified to \SI{10}{W} to excite the \SI{61.79}{MHz} Larmor resonance in  the selected NMR probes. 
The pulse width ($\tau \approx 7 \mu s$) is sufficiently short that its Fourier Lorentzian width ($1/ \tau $) overlaps the Larmor resonance  with enough power to excite the resonance. 
Using this same clock, the mixer down-converts the amplified and returning NMR waveforms to about 50 kHz for digitization by the waveform digitizers (WFDs).
\SI{61.74}{MHz} is then added back in the analysis.
Twenty pulser-mixer-multiplexer combinations (24 channels per multiplexer) are used to read out all 378 NMR probes of the \fpa.
In the following sections, we present details of the individual components.
\subsection{The frequency standard and frequency synthesizer}
\label{subsec:FreqStandard}
For all measurement systems in E989, RF signals are derived from a common 10-MHz  GPS disciplined frequency standard (SRS FS725/1C) that provides a continuous-wave sinusoidal signal. 
For the magnetometer system described here it provides the  reference standard for the frequency synthesizer and the sample clock for the waveform digitizers.
The \SI{61.74}{\mega\hertz} signal used to form the RF pulse to excite the FID and for mixing the FID signal down to the digitization frequency is generated by a frequency synthesizer (SRS SG 382). 
Video distribution amplifiers distribute this signal to the 20 pulser-mixer modules.
 
\subsection{Pulser-Mixer NIM module}\label{sec:pulsmix}
\begin{figure*}[ht]
    \includegraphics[width=1.0\linewidth]{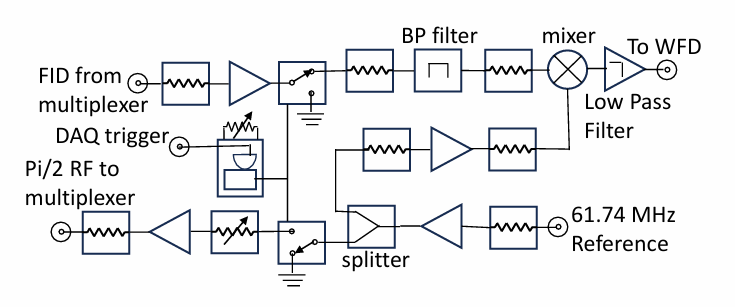}
    \caption{Block diagram showing pulser and mixer circuits using modular RF components. 
    Blocks consisting of fixed gain amplifiers (triangles) and attenuators (resistor symbols) provide signal gain and isolation. 
    In the lower right, the frequency reference is split between pulser (bottom) and mixer (top) signal paths.  
    Pulse width is determined by a one-shot multivibrator that  controls SPDT RF switches in the reference and returning signal paths. A switch (not shown) selects between the one-shot and its trigger pulse width.
    A voltage variable attenuator controls the output amplitude of the  pulse.
    FID signals are band pass filtered, mixed down to about 50 kHz and further amplified and filtered with a 4 stage low pass filter for the waveform digitizers (WFD).}
\label{fig:pulsblock} 
\end{figure*}

\noindent
The block diagram in figure~\ref{fig:pulsblock}
shows the RF signal processing architecture which uses active and passive modular RF components.
The lower branch corresponds to the pulser  and the upper branch to the mixer.
Common to both, the 10 dBm \SI{61.74}{\mega\hertz} reference signal is amplified with the combination of a \SI{10.5}{dB} amplifier (GVA-81+) and attenuator (LAT-6) for a net  power increase of \SI{4.5}{dB}.
A splitter (ADP-2-1W) provides \SI{11}{dBm} of power each for pulser and mixer branches of the diagram.

\subsubsection{Pulser}
\label{sec:PSF}
The pulser is used to form a combined control and excitation RF pulse of low power and adjustable length. 
A one-shot multivibrator (74HC221) (figure~\ref{fig:pulsblock}) triggered by a signal from the data acquisition computer (DAQ) generates an adjustable width TTL pulse in the range \SIrange{6}{16}{\micro\second} which controls a reflective single-pole-double-throw (SPDT) solid state switch (VSW2-33-10W+) connected to the \SI{61.74}{\mega\hertz} signal from the splitter. 
The total pulse length is determined by the time required to set up the RF switch-yard for high-power and the length of the RF excitation pulse $t_{\pi/2}$. 
This feeds the variable attenuator block consisting of a fixed attenuator (LAT-3) and voltage variable attenuator (MVA-2000+) with its control voltage range centered on a nominal value of \SI{-10.3}{dB}.
The subsequent gain block boosts the output power to \SI{10}{dBm} for transmission to the multiplexer.

\subsubsection{Mixer}
\label{sec:mix}
\noindent
The pre-amplified FID signal from the multiplexer (figure~\ref{fig:pulsblock}) has a maximum power of +\SI{1}{dBm} at the input to the mixer. 
This is increased to \SI{+15.3}{dBm} by the combination of an attenuator (LAT-1) and {\SI{15.3}{dB} amplifier (GVA-82).
The switch inhibits the returning FID during the RF pulse to avoid saturation of downstream signal-processing components. 
Signal bandwidth is limited with a band pass filter (BPF-B63+) to the range 63 $\pm$ 2}\unit{\mega\hertz}. 
The FID signal at the RF input of the mixer (ADE-1+, circle with X) has a maximum power of \SI{0}{dBm} which is \SI{15}{dBm} below the third-order intercept point (IP3).
Reference signal power is increased to \SI{17.5}{dBm} with a GVA-81 and LAT-4 combination and then attenuated by \SI{10}{dB} (LAT-10) to drive the LO input of the device at \SI{7}{dBm}.
The LO input of the mixer has a \num{2.5} voltage standing wave ratio (VSWR) and the \SI{10}{dB} attenuator provides isolation from the reflected signal. 
An active  low-pass filter section removes the sum frequency component  in the IF output signal and amplifies the approximately \SI{50}{\kilo\hertz} difference frequency for the WFDs. 
Four cascaded filter stages using operational amplifiers (Analog Devices ADA4898-1) make up the low-pass filter section.
The first stage is a first-order low-pass filter with a cutoff frequency of \SI{150}{\kilo\hertz} and a non-inverting gain of \num{10}. 
The next two stages are second-order low-pass filters (Sallen-Key topology) with cutoff frequencies at \SI{155}{\kilo\hertz} and \SI{177}{\kilo\hertz} respectively and Q-factors of \num{0.5}. 
These each have non-inverting gain factors of \num{1.2}. 
The final stage brings the nominal gain of the filter at \SI{50}{\kilo\hertz} to \num{25}, which is further adjusted to match the input voltage range of the WFDs. 

\subsection{Multiplexer Module}\label{sec:multiplexer}
\begin{figure}
    \includegraphics[width=.95\linewidth]{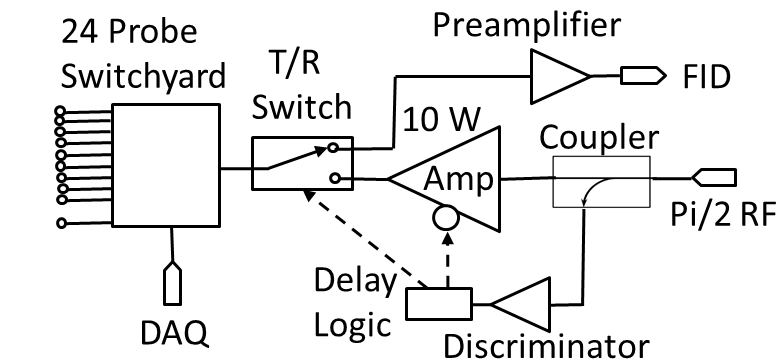} 
    \caption{Block diagram showing multiplexer components. The multiplexer serves as a switchyard for accessing individual NMR probes, increases pulse power to \SI{10}{\watt} and provides low noise pre-amplification of the returning FID signals. 
    The transmit-receive switch (T/R) determines the direction of power flow within the probe and is controlled by a network that prevents applying power until after the switch changes state.}
    \label{fig:muxblock} 
\end{figure}

\noindent
The multiplexer module, (block diagram shown in figure~\ref{fig:muxblock}), includes a switchyard for selecting one of \num{24} NMR probes and connecting it to the pulser or the mixer signal paths (T/R switch in the diagram), a low noise preamplifier to boost the FID signal for the mixer and a \SI{10}{\watt} pulse amplifier for exciting the probe. 
RF signal switching throughout the multiplexer uses single pole double throw (SPDT) reflective mode field-effect transistor (FET) switches (VSW2-33-10W+) rated for \SI{10}{\watt} continuous power, with \SI{0.5}{dB} insertion loss and \SI{40}{dB} isolation when switched off.

\subsubsection{Probe Switchyard}\label{sec:switchyard}
\begin{figure}
    \centering
    \includegraphics[width=0.95\linewidth]{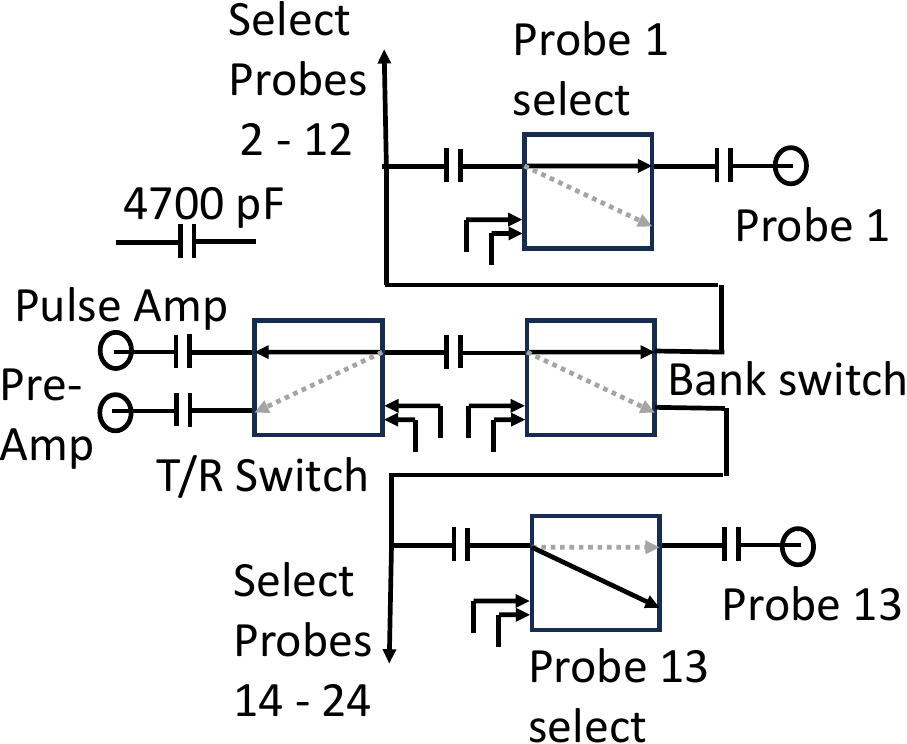}
    \caption{A partial schematic diagram of the probe switchyard and Transmit/Receive (T/R) switch in the multiplexer module. Individual probes are grouped in two banks of twelve probes with each probe selected through the setting of the bank switch and an individual probe select switch. The T/R switch determines the direction of power flow in the probes, either from the \SI{10}{Watt} amplifier or to the low noise preamplifier. Blocking capacitors are 4700 pF and switches are controlled by TTL signals and their complements. Circuit is shown selecting probe 1 for excitation by a $\pi$/2 pulse.}
    \label{fig:MultiplexerScheme}
\end{figure}

\noindent
The reflective mode probe switches connect a common signal bus between individual NMR probes when closed or present a high impedance when open. 
Figure~\ref{fig:MultiplexerScheme} is a section of the switchyard schematic diagram showing two banks of probe switches: the bank select switch, and the transmit-receive switch (see T/R figure~\ref{fig:muxblock}) that connects selected probes to either the pulse amplifier or preamplifier.
Grouping switches in two separate banks reduces signal loss from the combined input capacitance of all the switches.
The T/R switch is controlled in a coordinated way with the pulse amplifier to ensure switch transitions occur at low power. 
Section~\ref{sec:pulseamp} describes this in detail.
5-bit ASCII addresses from the DAQ's VME interface\footnote{Versa Module Eurocard system for digital inputs and outputs} encode the probe selection.  
Each of the five signals are filtered with an RC time constant of \SI{12}{\micro\second}. The signal line is further attached to an octal buffer (74HCT541). 
The two low order bits are input to six 2 line to 4 line decoders (74HC139) that select one of four probes in each of six groups of switches. 
The remaining higher order bits connect to one 3 line to 8 line decoder (74HC138), which selects one of the 6 groups.
An \textit{OR} of the first 3 outputs of the decoder (groups 1, 2, and 3) controls the switch selecting the probe bank. 
Front panel LEDs on the multiplexer show the selected probe channel. 

\subsubsection{Pulse Amplifier}\label{sec:pulseamp}
\noindent
The pulse amplifier (PA) increases RF  pulse power to \SI{10}{\watt}, achieving a \SI{90}{\degree} rotation of the proton spins away from vertical in about \SI{7}{\micro\second}. This is short compared to the \SI{100}{\micro\second} to  \SI{1000}{\micro\second} decoherence times $T_2^*$ of the probes, which are largely determined by the homogeneity of the magnetic field environment.
Actual pulse widths come from maximizing the amplitudes of FID waveforms.
Amplification proceeds in two stages, a driver amplifier and an output power amplifier. The output amplifier is an NPA1006 \footnote{MACOM https://www.macom.com/products/product-detail/NPA1006} GaN D-mode amplifier operating as a broadband linear amplifier with output power controlled through application of a gate-source bias voltage. 
A long-tailed pair, with one input controlled by a TTL signal, switches between the \SI{-5}{\volt} required for pinch off and an adjustable value between \SI{-5}{\volt} and \SI{-0.5}{\volt} for \SI{10}{\watt}-power operation. 
The output passes through two cascaded $\pi$ section low pass filters with \SI{115}{\mega\hertz} cutoffs, which is below the second harmonic of the \SI{61.74}{\mega\hertz} excitation frequency.
The driver amplifier is a TQP7M9105 \footnote{QORVO www.qorvo.com/products/p/TQP7M9105} with a power output of \SI{1}{\watt} and gain of \SI{17.9}{dB}. 
The NPA1006 and driver stage have a combined gain of \SI{28.9}{dB}.
To achieve \SI{10}{\watt}-operation, the \SI{10}{dBm} RF signal from the pulser (section~\ref{sec:PSF}) requires an additional \SI{10}{dB} of amplification to compensate for the \SI{7}{dB} signal loss incurred in the RG-174U coaxial cable used to connect the modules.
The input switch to the PA blocks the mixer's RF signal until full power bias has been applied to the output stage.
A DC-DC step-up converter, using an LT1930A (Linear Technologies) switching regulator, provides the \SI{+33}{\volt} drain voltage to achieve \SI{10}{\watt} output power from the NPA1006 amplifier. 
The VSW2-33-10W+ FET switches cannot switch between states while operating at their rated \SI{10}{\watt} power without incurring damage. 
The requirement of low power when switching is enforced through a combination of DAQ software and hardware logic in the multiplexer module. 
The DAQ controls the measurement sequence, setting  multiplexer probe select switches before enabling the $\pi /2$ pulse from pulser-mixer. 
\subsubsection{Pulse Amplifier Delay logic}\label{sec:logic}
\noindent
In the multiplexer module, biasing on the PA is delayed from the start of the $\pi /2$ pulse for the T/R switch to connect to the PA output at low power. 
This delay logic is shown in figure \ref{fig:muxblock}. 
Here the incoming RF signal from the pulser-mixer is split between the PA's normally open input switch and a discriminator (Linear Technologies LT1711) .  
When the amplitude of the incoming $\pi /2$ pulse exceeds threshold, the discriminator toggles a flip-flop (74HC74) at the reference RF frequency, which then continuously triggers a re-triggerable one-shot (74HC123) with delay set longer than twice the RF signal period. 
While triggered, it sets the TR switch to the output of the idling PA and biases the PA's output stage to achieve full power.  
The positive transition edge of this 74HC123 triggers a second 74HC123 which, after it times out, clocks a flip-flop (74HC74) closing the switch at the input of the PA to initiate the \SI{10}{\watt} $\pi$/2 pulse to the probe switchyard.
After the incoming $\pi /2$ pulse from the pulser-mixer module is completed,
the first 74HC123 times out biasing off the PA, connecting the TR switch back to the preamplifier and opening the input switch to the PA.
The resulting amplified $\pi /2$ pulse is about a microsecond shorter than that from the pulser-mixer, but the width is still correctly determined by maximizing FID amplitudes.

\subsubsection{Preamplifier}\label{sec:switches}
\noindent
Amplification of the microvolt-level FID waveforms returning from probes proceeds in two stages. 
The first stage consists of a limiter (RLM-33+) followed by an ultra low-noise amplifier (PSA4-5043+), operated at \SI{5}{\volt}, and mounted on the same circuit board as the probe switchyard. 
This first amplifier has a noise figure of \SI{0.73}{dB} and a gain of \SI{25.4}{dB}.
The limiter protects the low noise amplifier input from fast transients associated with \SI{10}{\watt}-excitation pulses.  
The second stage amplifier (Z-KL-1R5) is operated at \SI{12}{\volt} and located off-board.
It has a gain of \SI{41}{dB}, making the total gain from both stages \SI{66.4}{dB}. This second stage drives the \SI{50}{\ohm}-coaxial cable connection to the pulser-mixer module.

\subsection{Digitizer}
\label{subsec:Digitizer}
The amplified and down mixed FID signal is digitized with a Struck SIS3316 
ADC, where sampling is driven by the \SI{10}{MHz} reference clock. 
The digitizer has 16 channels and 16 bit resolution. Software down samples the data to \SI{1}{MHz}.
\subsection{System controller}\label{sec:fpga}
\noindent
The interface between the data acquisition computer and the electronic modules is provided by VME standard modules. 
These include a  computer to VME crate controller and waveform digitizers (WFD) from Struck Innovative Systems and an Acromag carrier board with mezzanine modules IP470A and IP-EP201 for digital input and output (DIO) signals. 
Three IP470A modules configured as outputs send addresses to multiplexers for probe selection, and one IP-EP201 with field programmable gate array (FPGA) processor Alterra Cyclone II along with code in the DAQ computer executes the rest of the measurement sequence. 
The IP-EP201 provides a hardware equivalent environment with \unit{\micro\second} timing precision to avoid latencies inherent in direct software control of the electronics. 
FPGA programs are written in VHDL\footnote{C like hardware description language, Institute of Electrical and Electronic Engineers IEEE Std 1076}, compiled and synthesized using Alterra's Quartus II development system and compiled code is uploaded to onboard memory via a JTAG\footnote{Industry standard hardware connector specification, Institute of Electrical and Electronic Engineers IEEE Std 149.1} port.
Within the EP-IP201 environment, individual logic components requiring time synchronization are implemented as processes in the VHDL code, initiated by common clock transitions. 
These include cascades of logic gates, derived clocks, counters, registers, and the state machine that generates the sequence of steps to control the hardware modules previously described.
\subsection{Operation modes}\label{sec:operation}
\begin{figure}[ht]
    \includegraphics[width=1.0\linewidth]{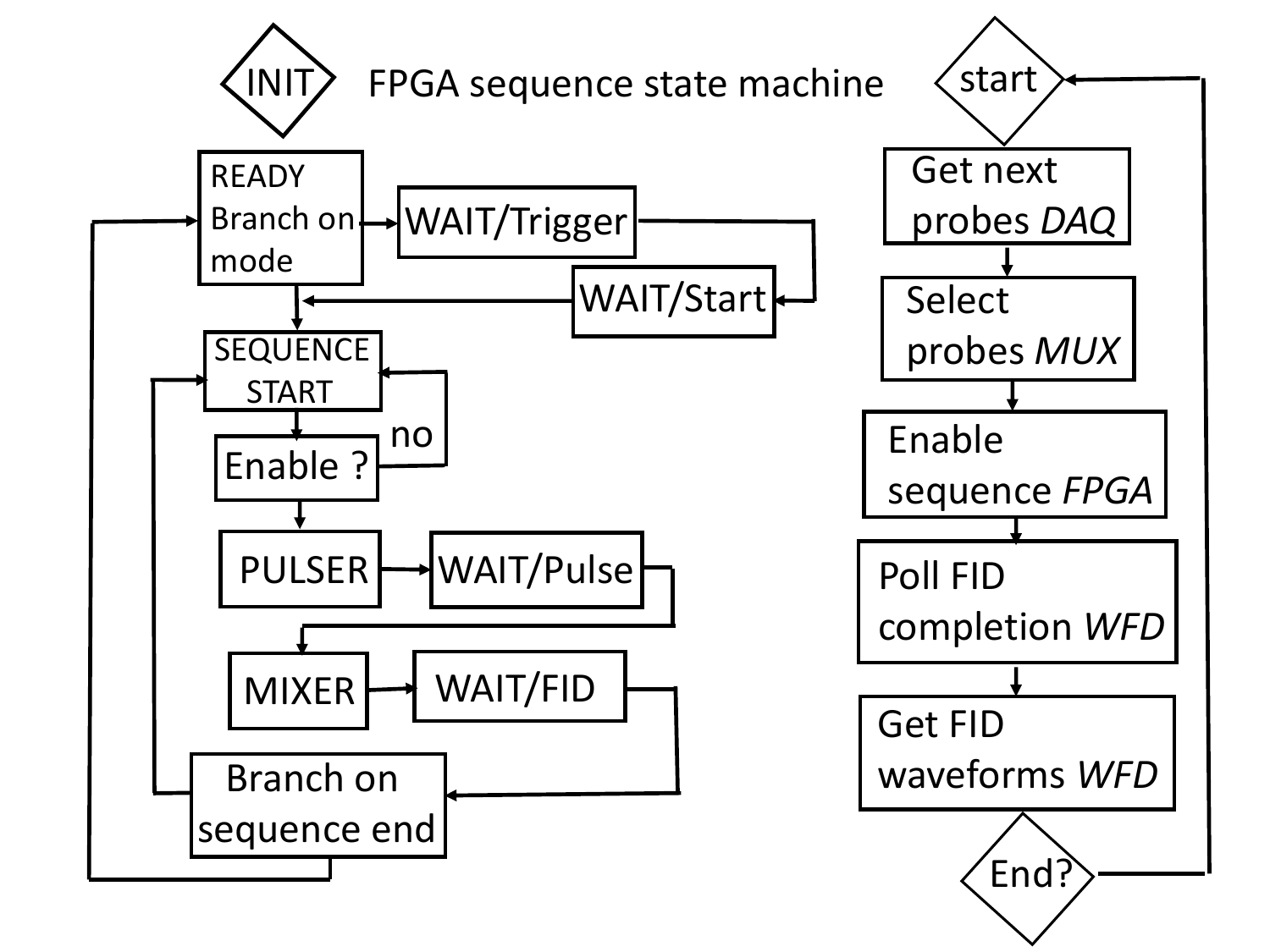} 
    \caption{The block diagram on the left shows the Moore State Machine architecture. Rectangular blocks represent individual machine states, and lines with arrows show the program sequence among the states. On the right is the corresponding diagram for code running in the DAQ computer. Together they provide control sequences to execute field measurements.}
    \label{fig:statemach} 
\end{figure}

Muon injection occurs with a \SIrange{1.33}{1.40}{\second} period preceded by a trigger at the start of each injection super cycle when muons are produced in the up-stream beam line. 
\SI{1.2}{\second} is sufficient for interrogating all \num{378} fixed NMR probes.
Three operating modes provide different options for the measurement sequence, including running asynchronously or synchronized with muon injection. 
In mode 1 the system runs asynchronously, taking measurements from all probes based on an internal timer set to \SI{1.4}{\second} or less. 
Muon g-2 production data was generally taken in mode 1.
In modes 2 and 3, measurements are synchronized with the start of the injection cycle whenever muons are present. 
The super cycle timeline has two groups of eight injected muon bunches with \SI{10}{\milli\second} spacing between bunches in each group.
In mode 2, the system makes five measurements per cycle: two coinciding with an injected bunch in each of the two groups, and three additional measurements uniformly filling out the cycle. 
In subsequent cycles, the start of the measurement is shifted to the next bunch in each group. To complete the measurement over all eight bunches, eight super cycles are required. 
Mode 3 is essentially mode 1, with measurements synchronized to the start of each muon super cycle and an additional configurable delay. 
In order to stabilize the field against drift through feedback with the magnet's power supply, continuous periodic field measurements need to be made, whether or not muons are present in the ring.
In modes 2 or 3, when the accelerator stops injecting muons and providing triggers, measurements continue triggered by the internal timer, which was set to the injection cycle period (see Tab.~\ref{tab:reg} register $34$).
A second counter, reset by accelerator triggers when present, provides a proxy for phase within the muon injection cycle. 
A 10 bit shift register ring counter (SR) provides the clock for the measurement timer with a nominal rate equal to its trigger rate divided by $10$. 
This can be varied in frequency by advancing or retarding which SR bit provides the clock's output after each SR tic, increasing or decreasing  the clock's frequency by 10 percent of its nominal value. 
When accelerator triggers become available while the system is running in modes 2 or 3, the clock's frequency is \textit{blue-shifted} (increased) or \textit{red-shifted} (decreased) until the measurement timer's phase coincides with the injection cycle phase and measurements proceed triggered externally.
Bit 8 in the status register (Tab.~\ref{tab:reg} register $2e$) indicates whether the source of measurement sequence timing is external or internal.
The Moore state machine architecture is shown on the left side in figure~\ref{fig:statemach}.
State changes occur on the rising edge of a \SI{1}{\mega\hertz} clock, which determines the timing precision within the sequence. 
The right side of the figure represents related steps in the data acquisition code.
Referring to figure~\ref{fig:statemach}, in a typical measurement scenario, probes are selected in the DAQ using dedicated digital ports and, to allow the measurement sequence to proceed, the enable bit is written to register $20$. 
Depending on the mode, the internal timer or external trigger starts the sequence, and after wait delays enters the pulser state which initiates $\pi$/2 pulses from the pulser electronics and triggers the WFDs to begin digitizing FID waveforms from the mixer electronics in the 20 NIM pulser-mixer modules. 
The mixer state then starts a timed pause of the sequence, which exceeds the time required for the DAQ cycle to complete. 
When the WFDs reach their preset measurement time, they signal the DAQ and FID waveforms are uploaded. The DAQ cycle then restarts, re-enabling the state machine. 
When the pause timer completes, the sequence restarts, providing a periodic time series of probe measurements at precise intervals. 

In addition to state machine and timer processes, the VHDL implementation includes VME accessible registers to set timer durations, enable the sequence and read timestamps. 
These are listed by their address offsets and whether they have read (R) or write (W) access in table~\ref{tab:reg}.
\begin{table}[ht]
\begin{center}
 \caption{VME accessible registers}
 \label{tab:reg}
  \begin{tabular}{ | c | l | }
    \hline
    Offset (hex) & Register Description \\ \hline \hline
    $02$ RW & configure DIO pins 1 - 16  \\ \hline
    $04$ RW & configure DIO pins 17 - 32  \\ \hline
    $20$ ~W & multiplexer ready  \\ \hline
    $22$ RW & start delay time  \\ \hline
    $24$ RW & bunch delay time \\ \hline
    $26$ RW &  FID delay time  \\ \hline
    $28$ RW & operating mode \\ \hline
    $2a$ R~ & FID timestamp low \\ \hline
    $2c$ R~ & FID timestamp high \\ \hline
    $2e$ RW & sequence status  \\ \hline
    $30$ RW & sequence parameters   \\ \hline
    $32$ R~ & $ T_{0}$ timestamp (external source)  \\ \hline
    $34$ RW & measurement period   \\ \hline
  \end{tabular}
\end{center}
\end{table}
The first two registers configure board pins as outputs for triggering pulser modules and WFDs, or inputs for external clock event signals.
After multiplexers are addressed, writing 1 to the register $20$ enables starting the FID sequence.  
Registers $22$ to $26$ configure the internal timers used in sequencing between machine states. 
Times are specified in units of \SI{10}{\micro\second}.
Register $28$ sets the operating mode (modes 1, 2, 3).
Registers $2a$ and $2c$ are the low order and high order 16 bits of the 32 bit measurement time stamp written to the data file.
Times are  in units of  microseconds elapsed since the previous external synchronization signal.
Register $2e$ gives the status of read sequences (current sequence ID). 
Register $30$ sets operating mode specific parameters. 
Register $32$ gives the internal clock time of a second external event signal input to the board.
Register $34$ sets the period of the internal measurement cycle timer.

%% file: 5_performance.tex
\section{Performance of the fixed-probe system}

\begin{figure}
\includegraphics[width=0.99\linewidth]{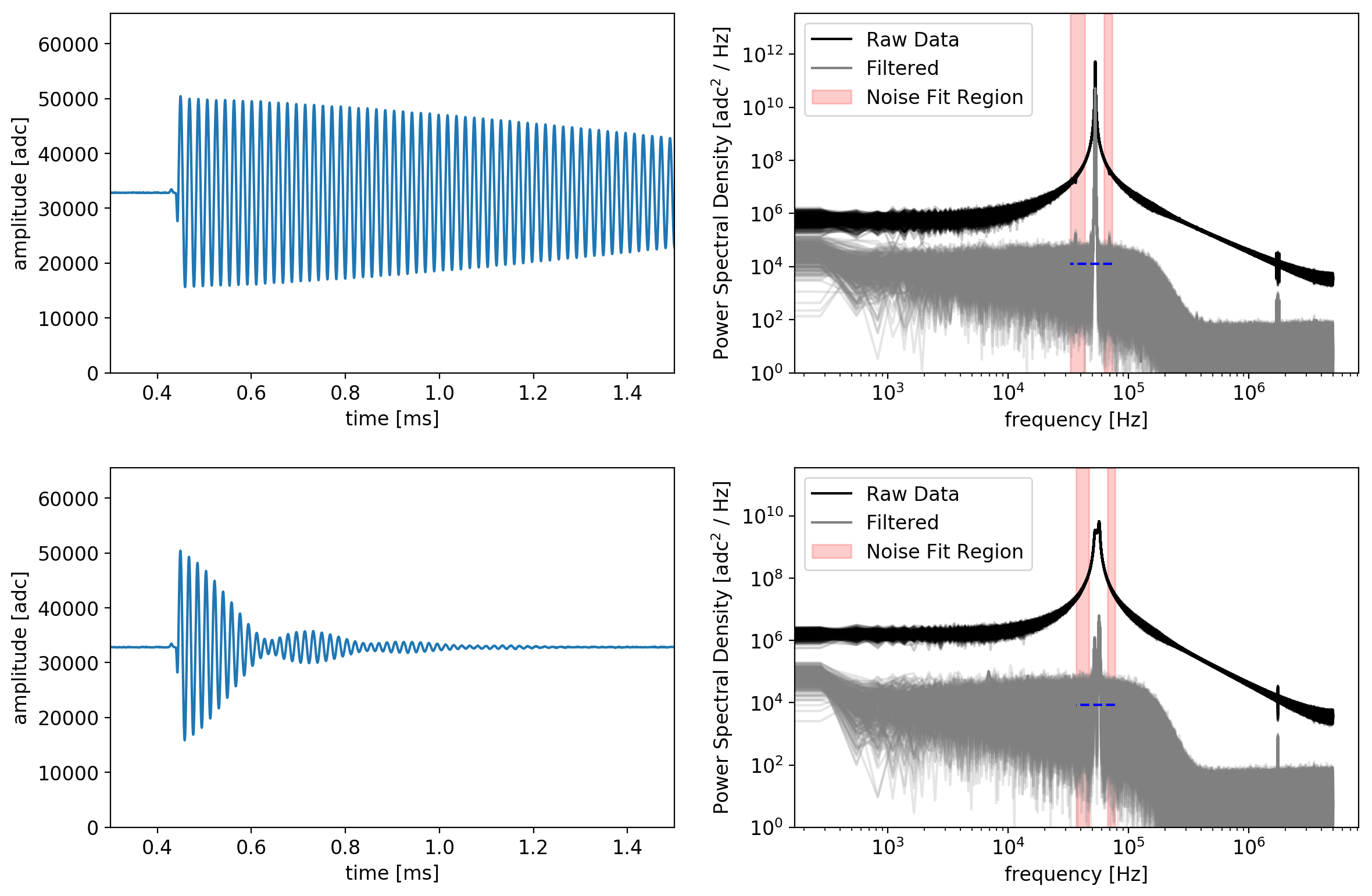}
   \caption{Free induction decay waveforms (FID, left) and power spectral density (PSD, right) of two fixed probe magnetometers. The probe shown in the upper panel experience only weak gradients, while the probe in the lower panel experiences stronger gradients, significantly shortening the FID. Their FID signals are characteristic of the two very different magnetic environments, with the rest of the probe waveforms lying somewhere in between. Power spectral density are shown for 310 different events from the same probe and for raw data, as well as a Hanning filter. The noise PSD under the peak is determined from the red shaded regions.}
   \label{fig:FID}
   \label{fig:performance_FFT}
\end{figure}

In this section we introduce performance metrics for the fixed probe magnetometer. These include
\begin{itemize}
    \item in situ measurement resolutions of individual NMR probes.
    \item a measure of probe performance independent of the local field environment.
    \item system component performance over the lifetime of the experiment.
    \item an example demonstrating the utility of good timing precision in the system controller.
\end{itemize}
Magnetometers, using petroleum jelly as proton-rich sample, are solid-state NMR probes. For this type of probes, gradients along the sensitive volume result in variations in the precession frequency within the probe's volume. 
The phases are initially synchronized at the $\pi/2$ pulse but decohere with time, which shortens the length of the FIDs.
Two FIDs in different magnetic field environments and thus different magnetic field gradients are shown in Figure~\ref{fig:FID}. The upper FID experiences weak external gradients and shows an exponential decaying envelope, while the lower FID experiences stronger gradients, significantly shortening and shaping the FID envelope. Zero-crossings of the envelope are typical for linear gradients in solid-state samples. 
In the right panels of Figure~\ref{fig:performance_FFT} the power spectral density (PSD) for these two probes are shown for raw data and using a Hanning-window for filtering, respectively. 
The PSD of the raw data can approximately be modeled by a sine oscillation with a free phase and exponential damping. 
The fixed probe magnetometers are sensitive to the signal's phase. While a plateau at frequencies below the reference frequency can be observed for cosine like functions, there is no plateau for pure sine like function. 
The raw signal PSD exhibits a second peak at \SI{1.74}{\giga\hertz} which is an aliasing of the \SI{61.74}{\giga\hertz} local oscillator and the sampling frequency of \SI{10}{\giga\hertz}. 
The filtered PSD isolates the resonance frequency from the white noise in the signal. The cut-off frequency of the low-pass filter section at $>\SI{100}{\mega\hertz}$ can be seen. Variations in the envelope only impact frequencies significantly below \SI{1}{\kilo\hertz}. 
It can be seen that the shorter FID has  contributions from a wide frequency range.

\begin{figure}
    \centering    
    \includegraphics[width=0.6\linewidth]{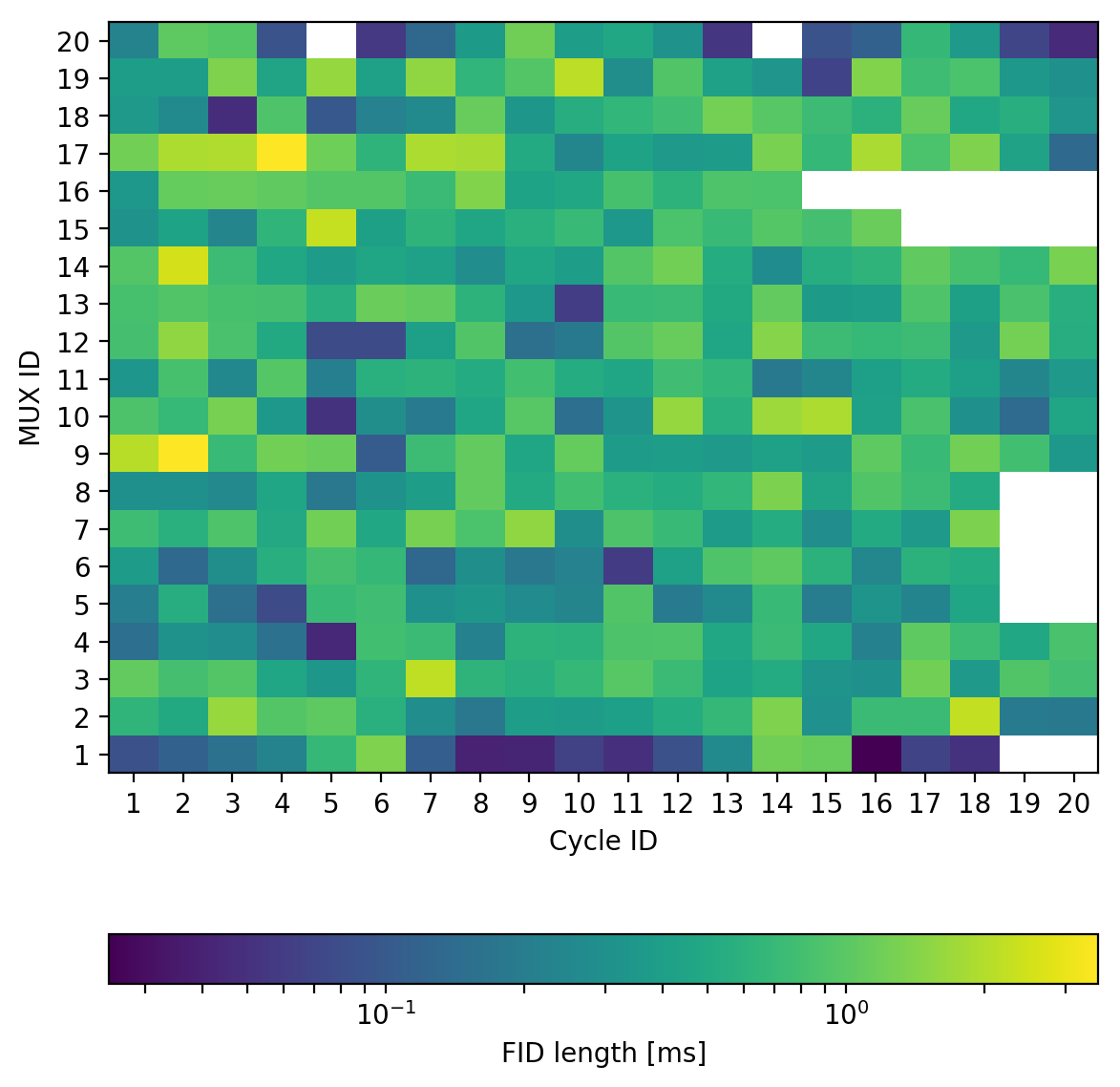}
    \caption{FID length for each probe indexed by MUX and readout cycle number (individual probe in the MUX). The FID length is the time in which the envelope of an FID drops to $1/e$ of the initial amplitude. The length of an FID is dominated by the strength of local gradients.}
    \label{fig:FID_length}
\end{figure}

\begin{figure}
    \centering
    \includegraphics[width=0.6\linewidth]{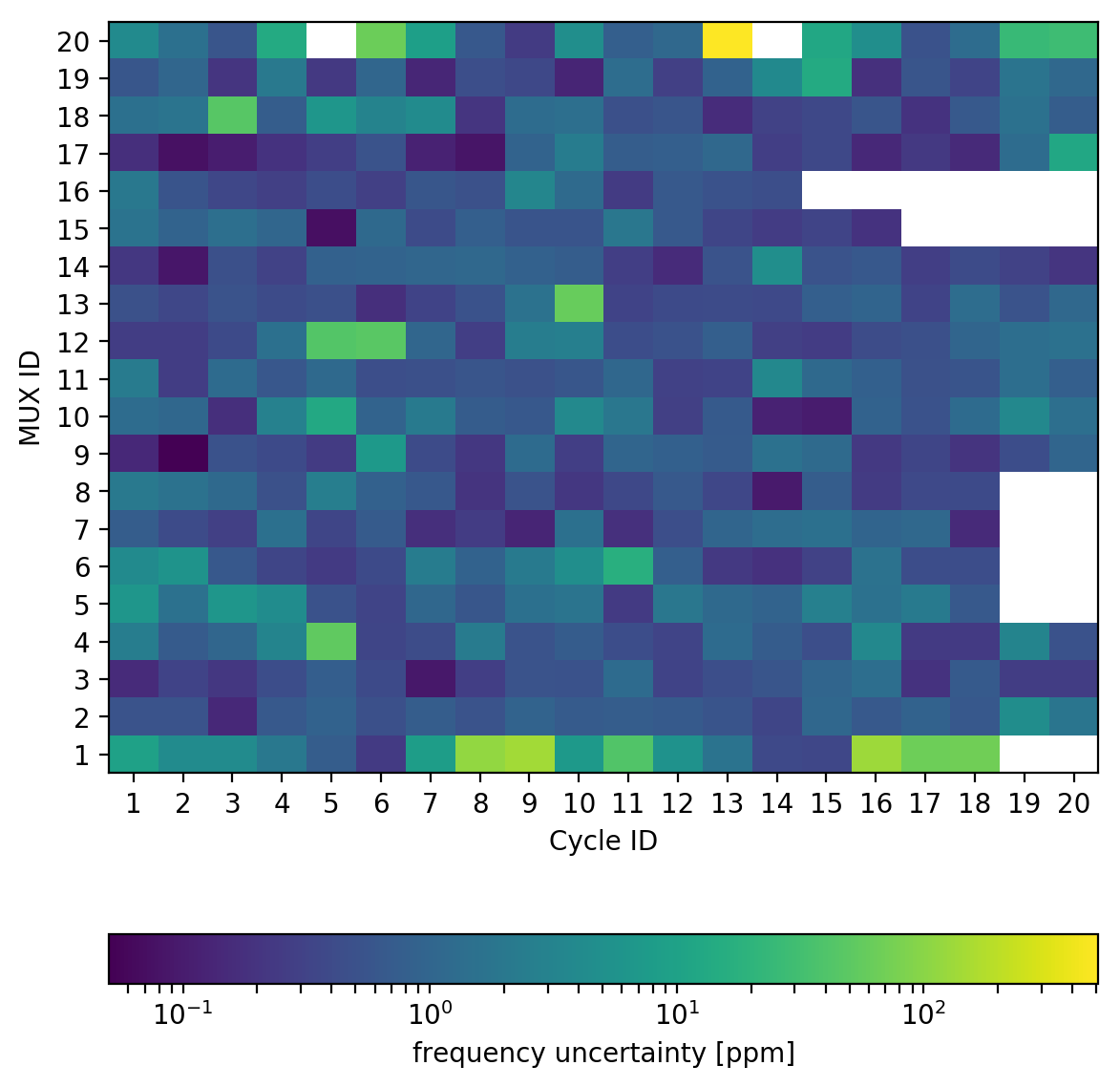}
    \caption{Single shot frequency determination uncertainty for each probe indexed by MUX and readout cycle number (individual probe in the MUX). 
    Due to different field gradient strengths in the magnetic environment of each probe, frequency uncertainties vary from \SI{150}{ppb} to \SI{13}{ppm} (central 90\%) with a median resolution of \SI{650}{ppb}. }
    \label{fig:freq_resolution_vs_MUX_and_cycle}
\end{figure}

\begin{figure}
    \centering
    \includegraphics[width=0.6\linewidth]{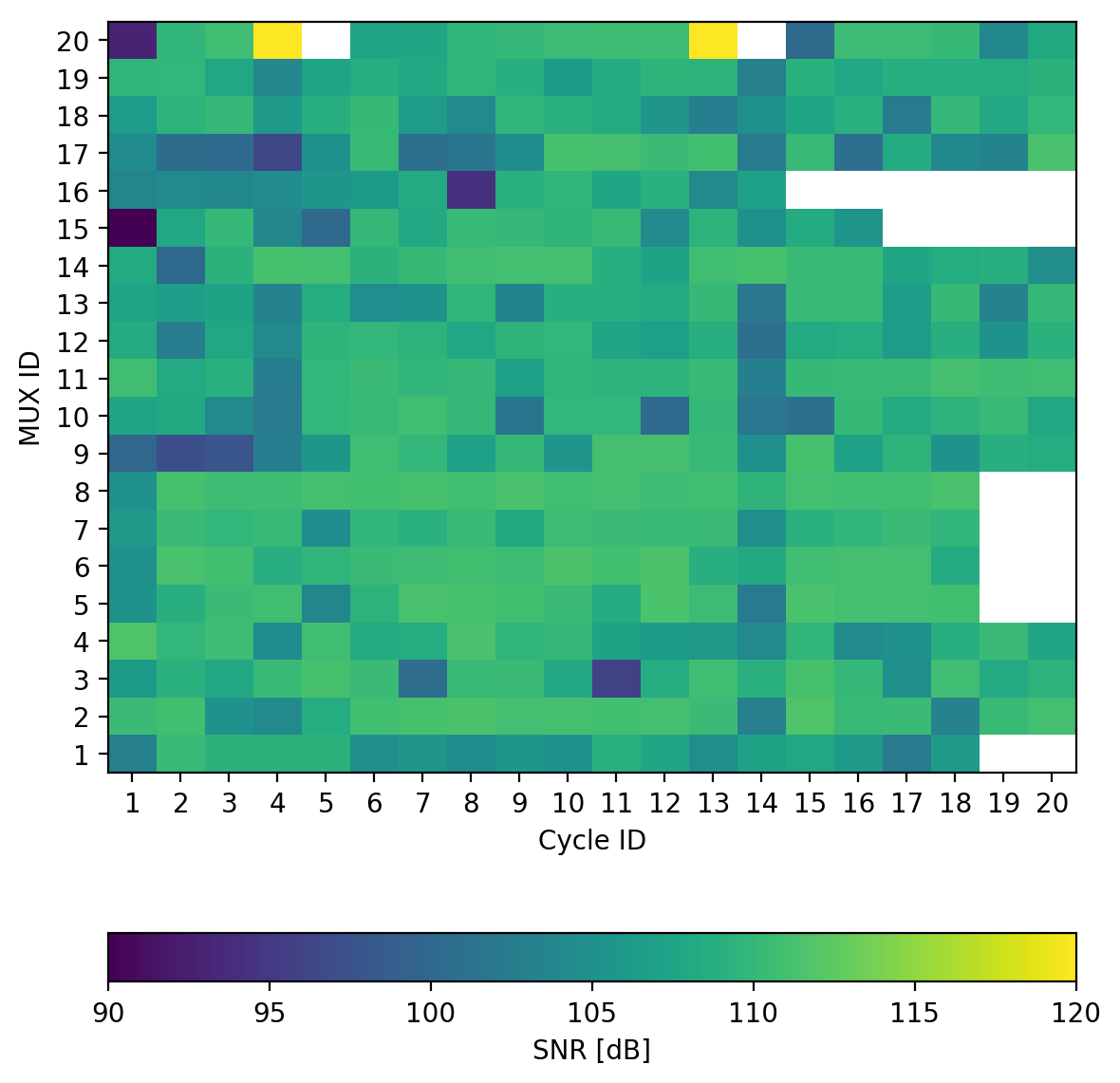}
    \caption{Signal to noise ratio (SNR) from maximal peak of power spectral density (PSD) spectra of each fixed probe indexed by MUX and readout cycle number (individual probe in the MUX). The SNR was calculated at the PSD peak position.}
    \label{fig:SNR_vs_MUX_and_cycle}
\end{figure}

Frequency resolution scales with the inverse of the  measurement time available for analysis. 
In analogy to the $T_2^*$ time defined in gaseous magnetometers, which exhibit a pure exponential decaying envelope, we generalize the definition of the FID length as the time that leads to a reduction of the initial amplitude to $1/e$, independent of the envelope shape. 
The FID length is shown in Fig.~\ref{fig:FID_length} for all probes. 
Magnetometer field measurements are given by proton precession frequencies obtained from FIDs. 
The phase and envelope functions are extracted using a Hilbert-transform. 
A static template is subtracted from the phase function to correct for constant local gradients. 
A linear fit is used to fit the phase function and extrapolated to the time of the $\pi/2$ pules~\cite{flowers_correcting_1995}. 
The full procedure is outlined in reference \cite{HONG2021107020}.
Frequency resolutions of all 378 probes are shown in figure~\ref{fig:freq_resolution_vs_MUX_and_cycle}. 
The frequency resolution are single shot frequency resolutions estimated from the phase fit function uncertainty. 
The vertical axis is the multiplexer (MUX) connected to the probe and the horizontal axis is the read cycle of all MUXs. 
To smoothen out field variations along the axis of the probes, shims were added to pole faces in the vicinity of the NMR probes.
While this effort was largely successful, these shims were incorrectly positioned around the inflector vacuum chamber, where muons are injected, and significant field gradients remained.
The inflector region, known to have large field gradients, has probes that connect to multiplexers MUX 1 and MUX 20. 
The poor resolution outliers are clearly visible at the top and bottom of the figure. Multiplexers' locations are roughly symmetric in storage ring azimuth. 
The darker horizontal band in the center of the figure corresponds to the opposite side of the ring from the inflector region and indicates larger field gradients there as well.
Probe resolutions in the fixed probe magnetic field environment range from \SI{150}{ppb} to \SI{13}{ppm} (central 90\%) with a median resolution of \SI{650}{ppb}.
Nearly all 378 NMR probes were used to determine the azimuthally averaged magnetic field for the muon g-2 experiment, only excluding a few probes close to the inflector that were too noisy.
Frequency resolution, while the important metric, depends on the local magnetic field homogeneity achieved by the shimming process. 
A measure of intrinsic performance less dependent on the local environment is the probe's signal-to-noise ratio (SNR). 
Following Gemmel {\it et. al}~\cite{gemmel_ultra-sensitive_2010}, we estimate the SNR from the averaged initial time-domain signal amplitude and the noise. 
We determine the noise power from the PSD of the filtered signal from two \SI{10}{\kilo\hertz} wide bands that are located \SI{10}{\kilo\hertz} above and below the signal peak.
The signal amplitude is the average of the FID envelope function before it decays to $1/e$. 
Individual probe SNRs are given in figure~\ref{fig:SNR_vs_MUX_and_cycle} by multiplexer and readout cycle number. 
Data was taken during the last run of the experiment and
probe SNRs continue to range between $\SI{90}{dB}$ and $\SI{120}{dB}$ with some outliers from probes in large field gradient regions. 
The large number of probes provided good spacial resolution over the storage ring azimuth.

\begin{figure}
    \centering
    \includegraphics[width=0.6\linewidth]{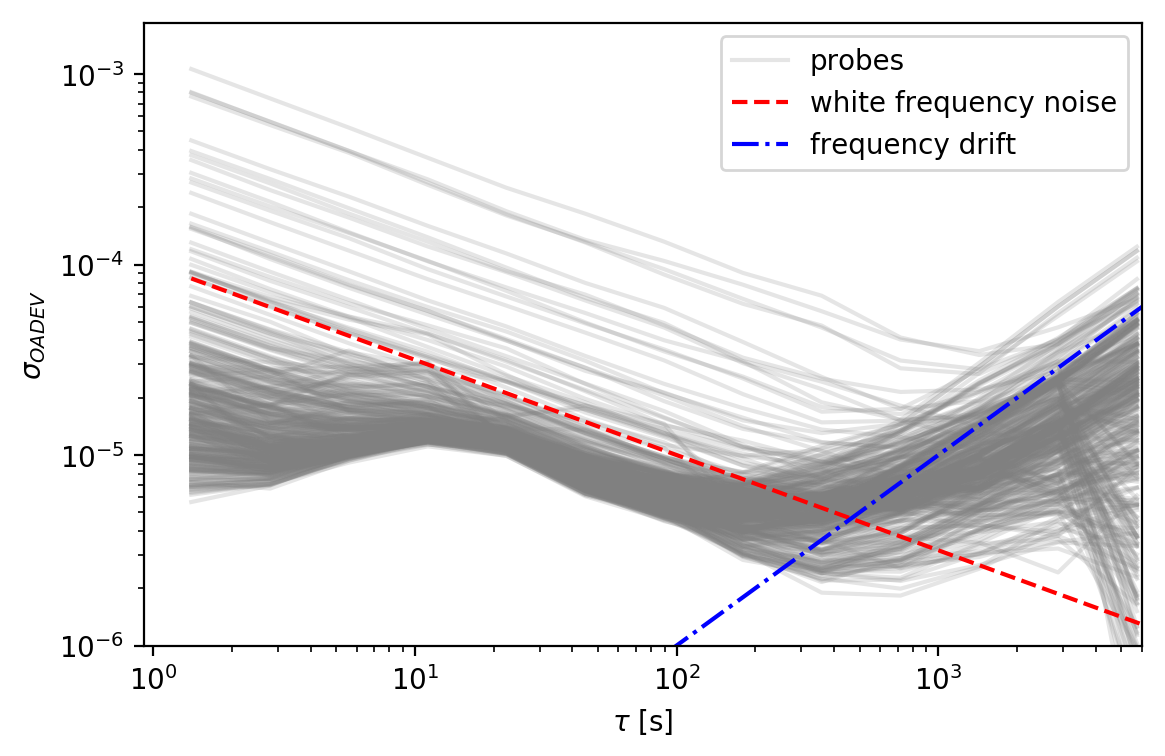}
    \caption{Overlapping Allan deviation as function of averaging time for each fixed probe magnetometer. The scaling for white frequency noise and frequency drift are shown in red and blue lines, respectively.}
    \label{fig:allan_dev}
\end{figure}

Another way to quantify the performance of the fixed probe magnetometers is the Allan deviation. We calculate the overlapping Allan deviation for a $\sim\SI{3}{\hour}$ long time series. 
Figure.~\ref{fig:allan_dev} shows the overlapping Allen deviation as function of averaging time for all probes. 
For time-scales significant lower then \SI{100}{\second} the measurements are dominated by white frequency noise, while above \SI{300}{\second} the measurements are dominated by frequency drift. 
A known oscillation on the time scale of $\sim\SI{2}{\minute}$ is observed and compensated by a feedback system on the main power supply. 
The fixed probe magnetometers are used to track this frequency drift between two trolley runs which are performed every few days.

Contributing to the excellent performance and longevity, electrical connections within the series resonant circuit that surrounds the probe's proton sample were either crimped or soldered, resulting in a low overall resistance and quality factor $Q =\num{60}$.
The comparable Q of those BNL E821 probes still operational in 2013 was \num{30}. 
The resonant amplification factor of the circuit scales with Q on resonance and contributes to the improved signal-to-noise ratio of E989 probes.
All fixed probes and probes in the measurement trolley continue to provide quality field measurements after six years. 
While this section emphasizes individual probes, the true performance measure of the magnetometer was demonstrated by magnetic field measurements in the E989 experiment; see for example references \cite{PhysRevA.103.042208}, \cite{PhysRevD.103.072002}.
In general, the fixed probe system surpassed its design requirements. In the latest published result of measurement campaigns 2 and 3, the overall magnetic field uncertainty was \SI{52}{ppb}~\cite{aguillard_detailed_2024} which is well below the technical design report goal of \SI{70}{ppb}~\cite{TDR2015} and the uncertainty attributed to magnetic field monitoring over time of \SI{17}{ppb} surpassed the design requirement of \SI{20}{ppb} as well.
The superferric storage ring magnet operates in non-persistence mode at a current around \SI{5170}{Amps}. Temperature variation in the experimental hall and pressure fluctuations in the cryogenic system result in magnetic field fluctuations. 
Based on fixed probe magnetometer measurements, a current feedback system in the magnet's power supply provides temporal stabilization of the magnetic field at the ppm scale. Its excellent performance is further detailed in~\cite{ops_paper}.
\begin{figure}
    \centering
    \includegraphics[width=0.95\linewidth]{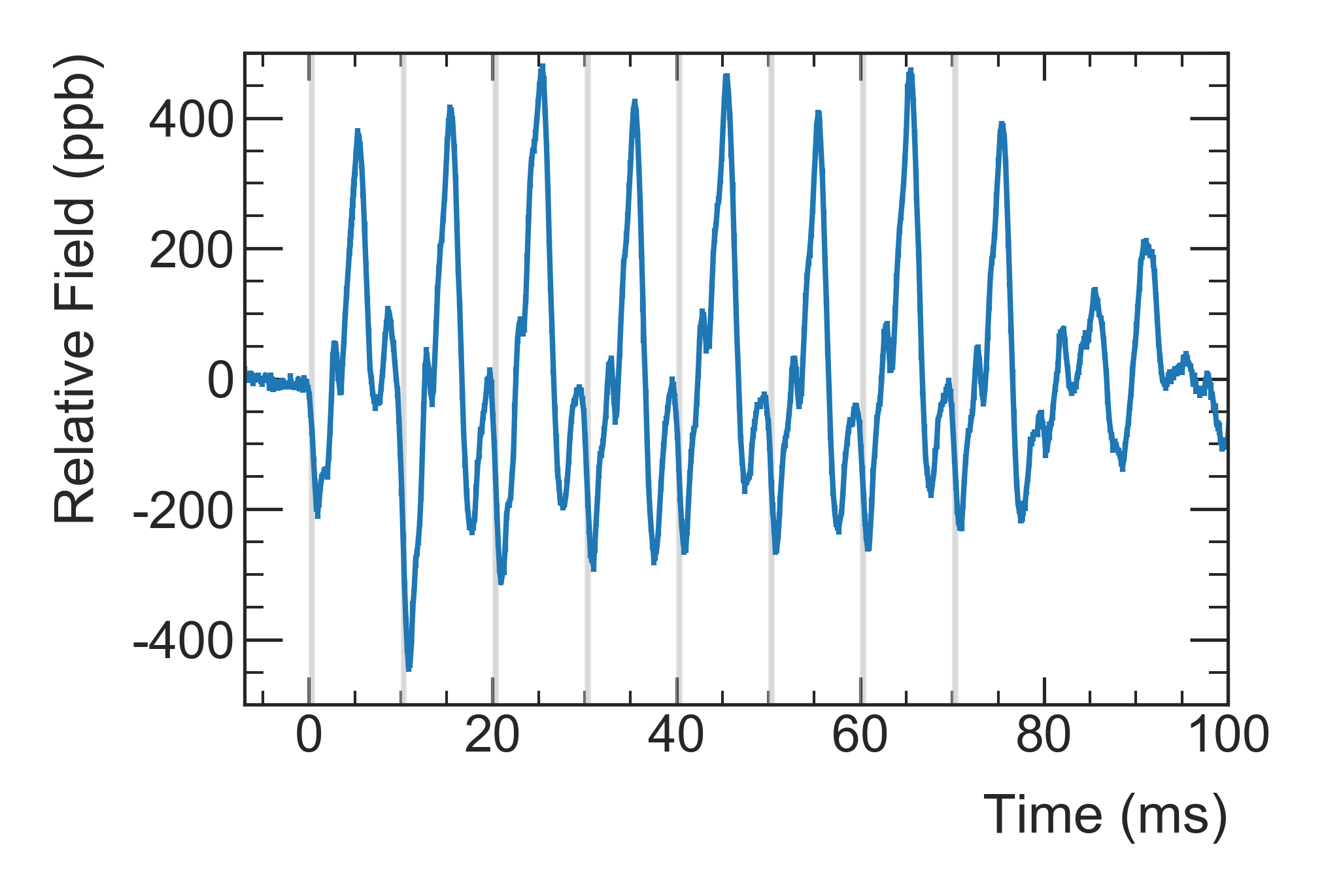}
    \caption{Composite plot of the transient magnetic field measured at the center of electrostatic quadrupole Q3L. The persistent field transient is sampled by continuously increasing measurement start time delays at each subsequent muon injection cycle. The plot shows the time series of these samples. Vertical lines are the times of muon bunches, when electrostatic quadrupoles are pulsed on. Figure from~\cite{PhysRevA.103.042208}.} 
    \label{fig:transient}
\end{figure}

\subsection{Electrostatic Quadrupole Transients}

When measuring asynchronously (mode 1), fixed probes located in vacuum chambers containing electrostatic quadrupoles (ESQ) showed larger RMS variations than probes in vacuum chambers without them.  
Muons are constrained vertically in the storage ring by these electrostatic quadrupoles, which have to be uncharged after each muon fill, to prevent electron trapping in a quasi-Penning trap and build up of plasma. 
Thus the electrostatic quadrupoles are charged and discharged corresponding to the muon beam injection. 
The beam is delivered in two groups with a series of eight pulses spaced by \SI{10}{\milli\second} per group. 
The second group occurs \SI{266.7}{\milli\second} after the first series and the entire structure repeats every $\sim\SI{1.4}{\second}$. 
The electrostatic quadrupoles are charged with a rise time of~\SI{5}{\micro\second} then held charged for \SI{700}{\micro\second} and afterwards uncharged \cite{muon_g2_collaboration_beam_2021}. 
Synchronizing measurements with these bunches showed previously unknown fast magnetic field transients. 
This pulsing leads to mechanical vibrations induced by Lorentz forces in quadrupole electrodes, generating fast magnetic transients by eddy currents at mechanical vibration frequencies which are much slower then the quasi-instantaneous pulse. 
The average storage ring magnetic field is measured every $\SI{1.4}{\second}$. While muons occupy the storage volume less than \SI{1}{\percent} of that time they do experience these transient fields. 
The $\SI{1}{\us}$ precision of the sequencer programmed in the IP-EP201 FPGA allowed precision time scanning of field measurements across the duration of muon bunches, where the measurement start-time delay is increased at each injection cycle. 
High-frequency magnetic-field oscillations are strongly attenuated by the aluminum vacuum chamber walls and require a custom probe assembly inside the vacuum region to measure them.
When muons were not being stored in the ring, an NMR probe, sealed against vacuum by a PEEK housing, was placed at the center of the electrostatic quadrupole (near zero electric fields) and measurements were taken while the quadrupole were pulsing.
The time series of these scans shown in figure $\ref{fig:transient}$ is the aliased persistent magnetic transient waveform seen by muons measured with NMR precision. 
For the analysis of this transient about 20 FIDs are measured for each time delay and the first \SI{2}{\milli\second} of the \SI{4.096}{\milli\second} long FID are used to extract the precession frequency. 
The bandwidth of the probe is limited by the skin depth effect in the aluminium sleeve, where the aluminium sleeve thickness corresponds to the skin depth at \SI{27}{\kilo\hertz}.
The attenuation of the aluminium sleeve as function of frequency was measured with a special setup consisting of a pair of Helmholtz coils which apply an oscillating external field and a small pickup coil that fits in the aluminium sleeve. 
For frequencies $<\SI{100}{\hertz}$ the attenuation of the signal is negligible and drops to $\sim40\%$ for frequencies of \SI{10}{\kilo\hertz}. 

%% file: 6_conclusion.tex
\section{Summary}
The E989 muon g-2 experiment at Fermi National Accelerator Laboratory aims for a twofold reduction of systematic uncertainties in storage ring field measurements compared with BNL E821 requiring ppb level accuracy. 
The technique of pulsed proton NMR is capable of part per billion uncertainties and, following BNL, it was used here as well. 
The fixed probe magnetometer continually tracks the magnetic field whether muons are in the storage ring or not. 
NMR probes were completely redesigned with robust construction techniques and RF electronic modules were developed using modern modular electronics. 
All 378 NMR probes embedded in the ring vacuum chamber walls were operational, providing a uniform sampling of the magnetic field in azimuth. 
The initial systematic goals of the experiment were more than met by this magnetometer.

%% file: 7_acknowledgement.tex
\section*{Acknowledgments}
We thank Fermilab management for providing funding for magnetometer construction, the UW Physics Department machine shop for precision machining of parts for more than 450 NMR probes and the CENPA staff for help in navigating the procurement process and providing resources for establishing the \SI{1.45}{T} field for testing the system. 
We thank Dr. Simon Corrodi for sharing his analysis results in figure \ref{fig:transient}.
This work was supported by the DOE Offices of Nuclear Physics (DE-FG02-97ER41020). 
Prof. Dr Martin Fertl and Dr. René Reimann receive support from the Cluster of Excellence “Precision Physics, Fundamental Interactions, and Structure of Matter” (PRISMA+ EXC 2118/1) funded by the German Research Foundation (DFG) within the German Excellence Strategy (Project ID 39083149).

%% file: 8_Appendices.tex
\section{Appendices}
\label{sec:Appendices}
\subsection{Probe assembly process}
\label{subsec:ProbeAssemblyProcess}
\begin{figure}[ht]
    \centering
    \includegraphics[width=0.5\linewidth]{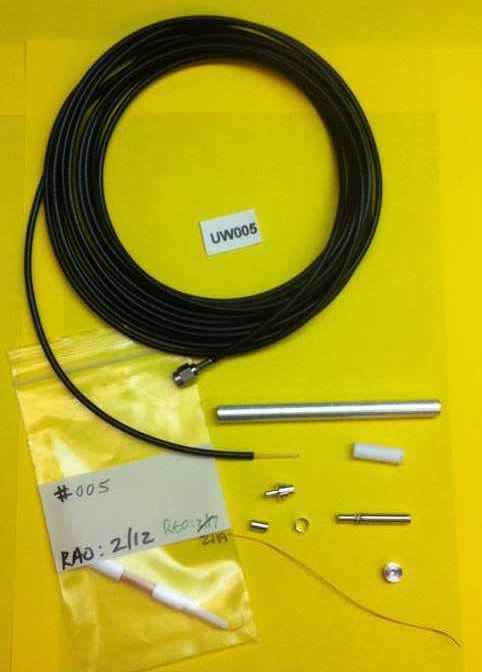}
    \caption{A typical assembly kit for a single NMR probe containing all required construction pieces. For quality control purposes, the individual assembler of the probe are indicated on the bag and each probe was assigned a unique identification number.}
    \label{fig:AssemblyKit}
\end{figure}

More than 450 pNMR probes were assembled efficiently and reproducibly. 
We developed a standard assembly procedure based on an \textsl{assembly kit} that contained all parts required to assemble an individual probe from start to finish. 
Such an assembly kit is shown in figure\,\ref{fig:AssemblyKit}.
In addition, we developed several mechanical tools to ensure precise assembly of the mechanically delicate parts. 
In the next subsections, we give examples of procedures used in the preparation of individual parts and sub-assemblies and present some of the special tools used in the assembly process. 

\subsubsection{Cleaning procedure for the PTFE parts}
\label{subsubsec:CleaningPTFE}
Machining fluid residues from all parts made from PTFE were removed with a three stage cleaning procedure:
\begin{itemize}
\item \SI{20}{\minute} of ultrasonic cleaning in pure acetone heated to \SI{60}{\degreeCelsius}. 
\item \SI{5}{\minute} of ultrasonic cleaning in pure 2-propanol heated to \SI{60}{\degreeCelsius}.
\item \SI{5}{\minute} of ultrasonic cleaning in deionized water followed by overnight drying at room temperature.
\end{itemize}
Any handling of parts was performed with gloved hands.
\subsubsection{Cleaning procedure for the aluminum parts}
\label{subsubsec:CleaningAlu}
Machining fluid residues from aluminum parts were removed with a similar three stage cleaning procedure:
\begin{itemize}
\item \SI{20}{\minute} of ultrasonic cleaning in pure 2-propanol heated to \SI{60}{\degreeCelsius}. 
\item Rinsing with deionized water.
\item \SI{5}{\minute} of ultrasonic cleaning in deionized water.
Removal of all remaining water with lint-free laboratory grade wipes, followed by overnight drying at room temperature.
\end{itemize}
Gloves were also required for handling these parts.
\subsubsection{Filling the \sh with petroleum jelly}
\label{subsubsec:FillingJelly}
\begin{figure}[htbp]
\centering
\includegraphics[width=0.5\linewidth]{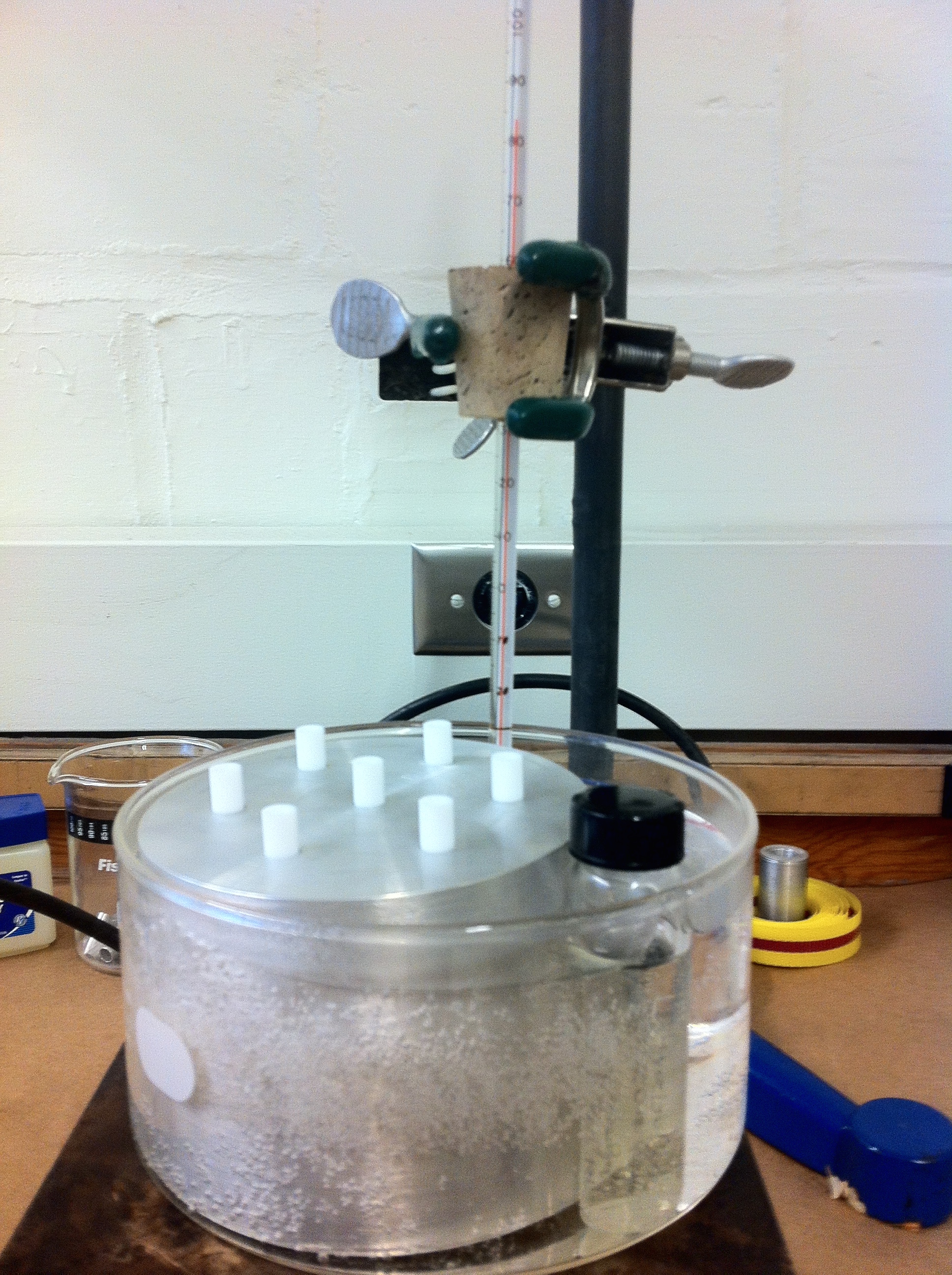}
\caption{This fill station was used to melt the petroleum jelly in a water bath on top of a heater plate, which was also used to pre-heat the \sh. A glass jar with molten petroleum jelly can be seen to the right of the aluminum block.}
\label{fig:PetroleumJellyFillStation}
\end{figure}

\begin{figure}[ht]
\centering
\includegraphics[width=0.5\linewidth]{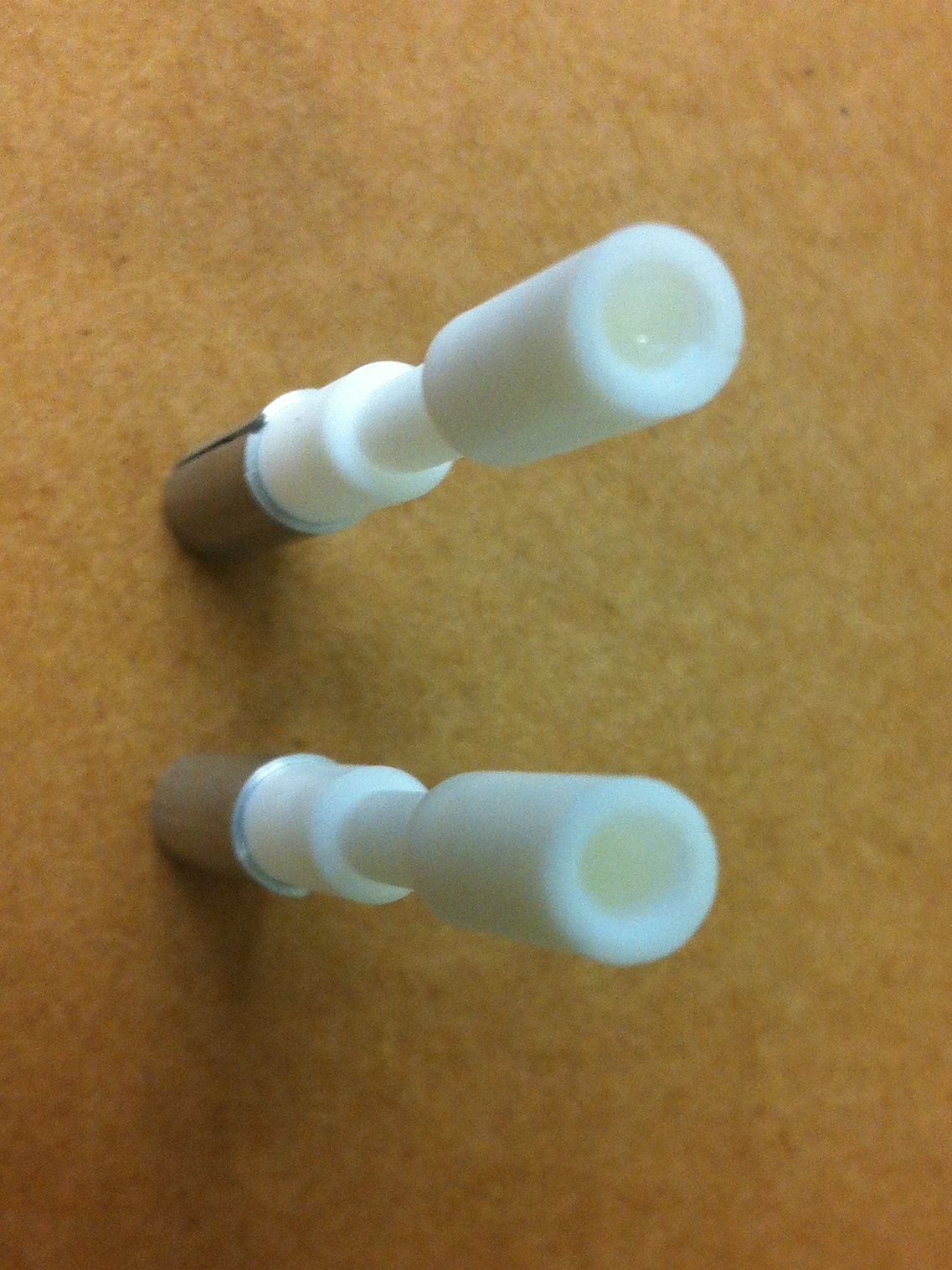}
\caption{Two sample holders shown with their sample cavities filled with petroleum jelly (slight yellow tint). Aluminum sleeves that protect the \sh during winding of the serial inductor coil are also depicted here.}
\label{fig:FilledSampleHolders}
\end{figure}

The petroleum jelly and \sh are pre-heated and maintained at about \SI{80}{\degreeCelsius} during the filling process in order to have enough handling time. 
The setup used to fill seven \sh cavities at a time is shown in figure\,\ref{fig:PetroleumJellyFillStation}.
The following procedure helped ensure that no air bubbles were introduced in the sample cavity
\begin{enumerate}
    \item \textbf{Warming the \sh}: Place the \sh holding block into the water bath and warm it up to \SI{90}{\degreeCelsius}.
    \item \textbf{Melting the petroleum jelly}: Melt the petroleum jelly in a glass jar by putting it into the hot water bath next to the Al block. Warm the petroleum jelly until its viscosity is low enough that it can be sucked into the syringe through a blunt stainless steel needle (gauge 16).
    \item \textbf{Preheating the glass syringe}: Preheat the glass syringe in a heated jacket regulated to \SI{85}{\degreeCelsius} to prevent the petroleum jelly in the glass syringe from becoming solid. Install the stainless steel needle with Luer lock onto the glass syringe.
    \item \textbf{Preheating the \sh pieces}: Load seven \sh pieces into the holes of the holes of the aluminum block and wait some minutes until all sample holders warmed up to the temperature of the Al block.
    \item \textbf{Transfer of petroleum jelly}: Take the preheated glass syringe and bring the needle into the glass jar with the molten petroleum jelly. Allow the needle to heat up for some seconds before you try to fill the syringe. Draw in about \SI{5}{\milli\liter} of the molten petroleum jelly. Stick the stainless steel needle into the sample volume of the \sh until it reaches the bottom of the cavity. Slowly inject petroleum jelly into the sample space while retracting the needle (This is important as just pulling the needle out causes bubbles to form). DO NOT overfill the sample space. Fill it up to the middle of the thread for the \icond.
    \item \textbf{Remove any air bubbles}: If during inspection of the sample space, air bubbles are detected, they can be removed by lowering the needle again to the bottom of the cavity and by injecting more petroleum jelly. Alternatively, one can suck the air bubble back into the needle/syringe. 
    The first way is highly recommended.
    \item \textbf{Cool down}: Remove the aluminum block from the water bath and let the whole block cool down to room temperature.
\end{enumerate}
Two filled sample holders are shown in figure\,\ref{fig:FilledSampleHolders}.
\subsubsection{Winding the serial inductor coil}
\label{subsubsec:CoilWinding}
\begin{figure}[ht]
    \centering
    \includegraphics[width=0.5\linewidth]{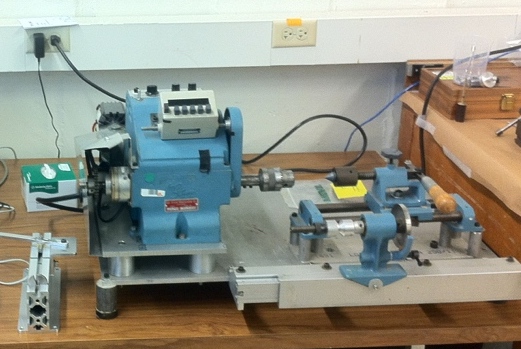}
    \includegraphics[width=0.5\linewidth]{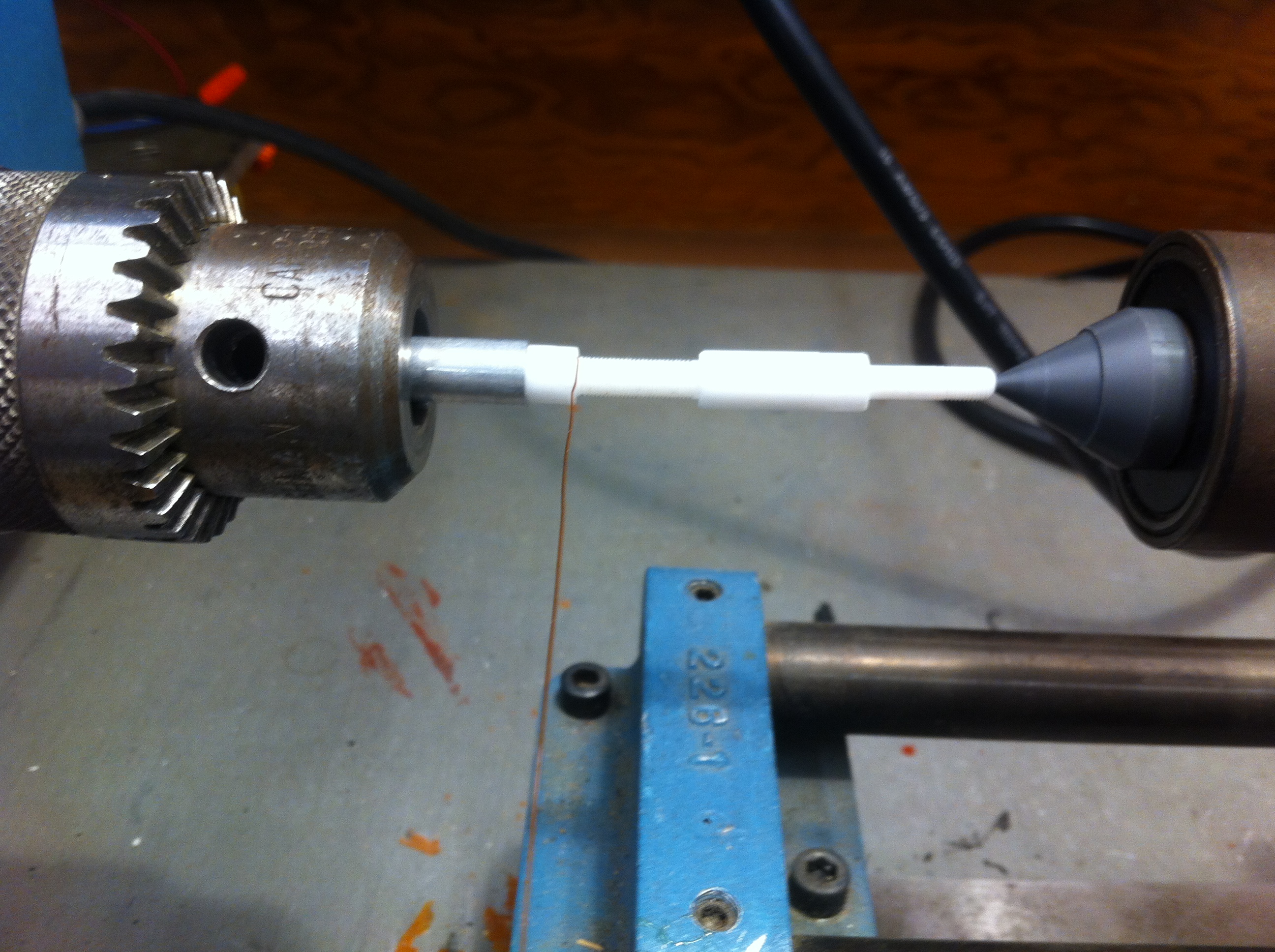}
    \caption{The series inductor coils were wound with a refurbished slowly rotating coil winding machine. The gray box on top of the blue motor box is a turn counter. Only non-ferromagnetic material is in physical contact with the PTFE sample holder.}
    \label{fig:SerialCoilWinding}
\end{figure}

\begin{figure}[ht]
    \centering
    \includegraphics[width=0.5\linewidth]{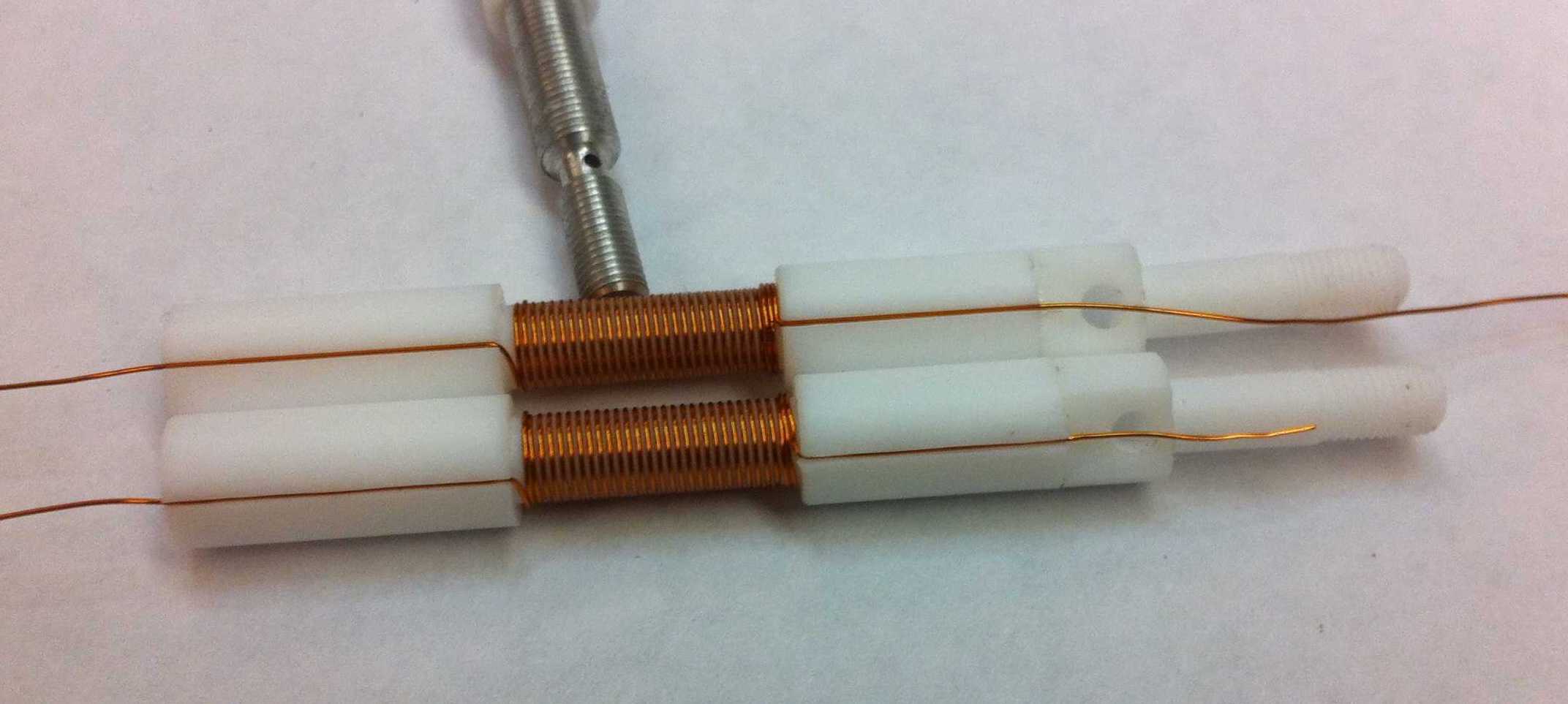}
    \caption{A set of two finished series inductor coils wound on their PTFE sample holder.}
    \label{fig:WoundSerialInductors}
\end{figure}

The wire of the inductor coil is first pressed into its dedicated groove in the \sh  (this protects the wire varnish from being scraped off by the aluminum sleeve installed in a later step) and the first three turns where wound manually. 
Coil windings are guided by external M4.5x0.5 threads in the bottom of the coil region. This allowed maintaining a constant tension on the wire during the winding process. 
After this, an aluminum sleeve (see figure\,\ref{fig:FilledSampleHolders}) was attached to protect the \sh from ferromagnetic impurities of the machine chuck on the coil winding machine.  
The coil winding machine shown in figure\,\ref{fig:SerialCoilWinding} was used to wind the remaining 29 coil turns into the threads. 
To avoid bending the \sh during the winding process a dead center made from plastic was used on the far end of the \sh (also shown in the figure). 
After all 32 turns have been wound the free wire end was secured in the second half of the dedicated wire groove. 
All part handling was performed with gloved hands to protect the clean parts. A set of two finished series inductor coils is shown in figure\,\ref{fig:WoundSerialInductors}.
\subsubsection{Soldering a wire loop on the inner conductor}
\label{subsubsec:SolderingAlu}
\begin{figure}
    \centering
    \includegraphics[width=0.5\linewidth]{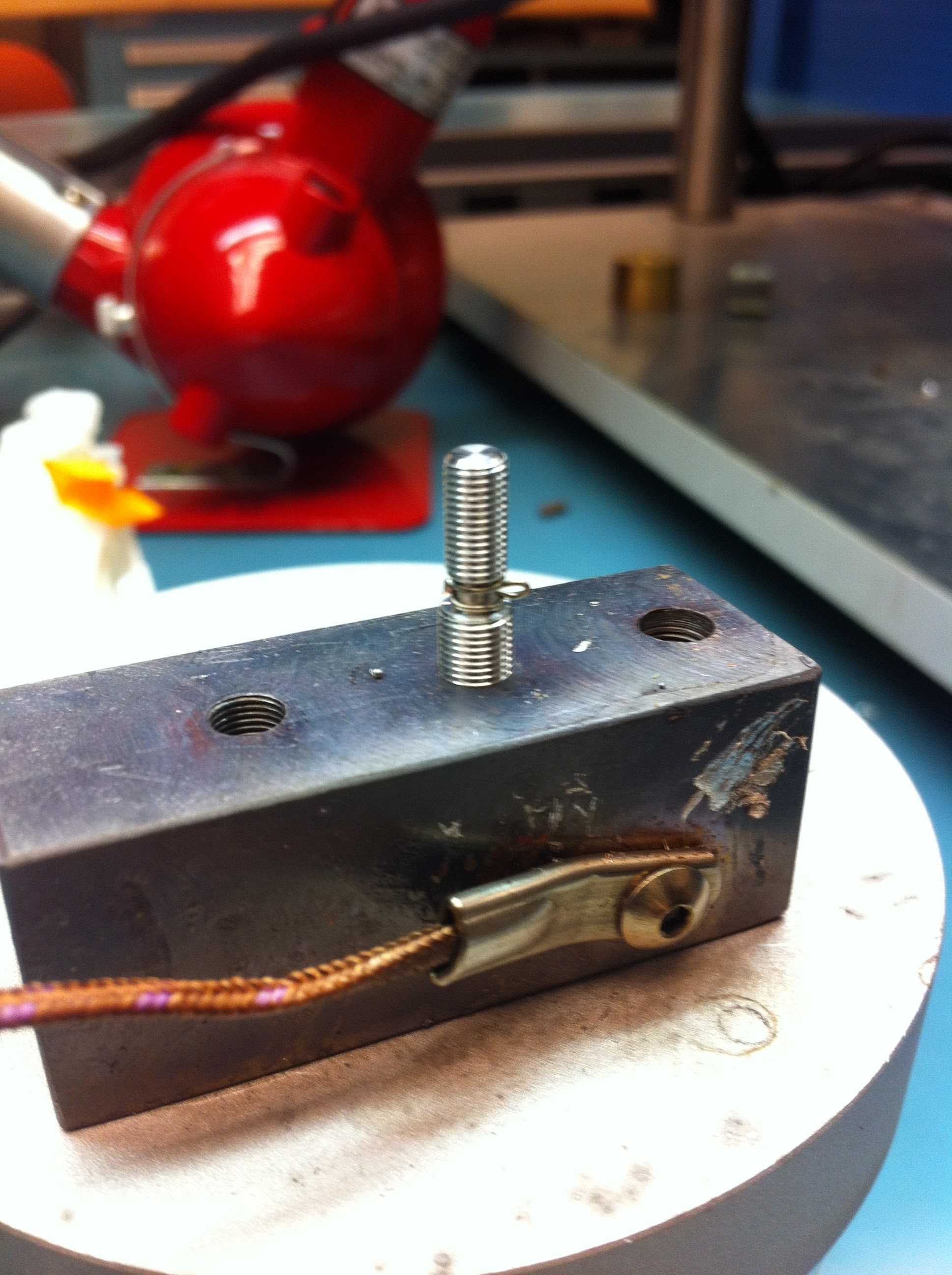}
    \caption{One inner conductor piece screwed into a copper block with threaded holes sitting on a hot plate to facilitate the ultrasonic soldering process.}
    \label{fig:InnerConductorInHotPlate}
\end{figure}

\begin{figure}[htbp]
\centering
\includegraphics[width=0.5\linewidth]{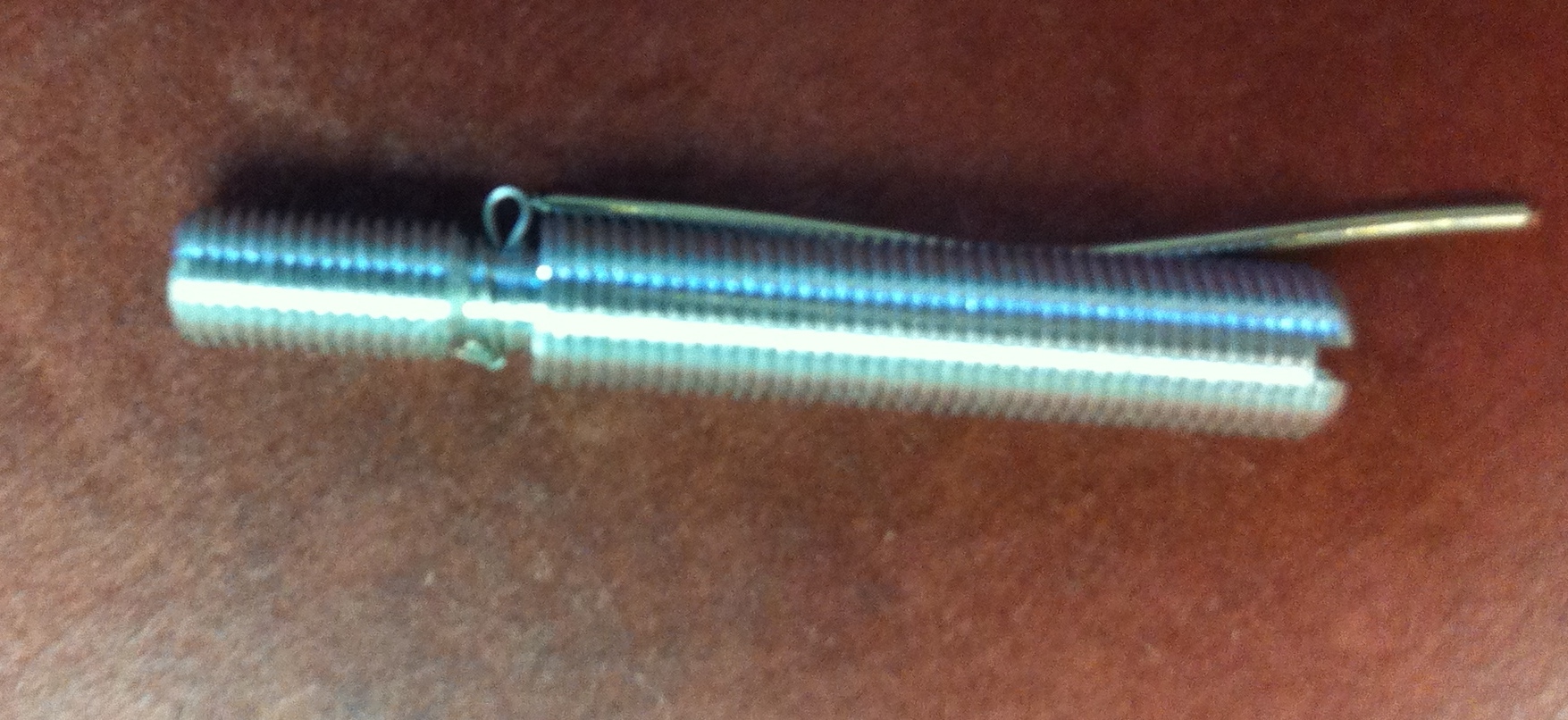}
\caption{Soldering the connection loop into the \icond is performed with an ultrasonic soldering iron.}
\label{fig::SolderingEye_2}
\end{figure}

The serial inductor coil is connected to the \icond in a two-step procedure. 
First a wire loop is soldered to the aluminum inner conductor using a  flux-less, ultrasonic soldering technique. 
This reduces the risk of long-term corrosion-induced contact failure. The power of the ultrasonic soldering iron is insufficient to heat the \icond pieces to the solder melting point so they were preheated by screwing them into a copper block sitting on a hot plate, see figure\,\ref{fig:InnerConductorInHotPlate}. 
In the second step, the series inductor coil is soldered to the wire loop.
The ultrasonic soldering procedure:
\begin{enumerate}
    \item \textbf{Forming of the loop}: A short (1 inch) piece of bare AWG 26 magnet wire is pulled through the \SI{1}{\milli\meter} diameter hole in the \icond such that a small loop is remaining on one side that prevents the wire from falling through the hole.
    \item \textbf{Securing the wire}: The wire is secured in the hole by bending the two wire ends in opposite directions around the \icond, as shown in figure\,\ref{fig::SolderingEye_2}.
    \item \textbf{Heating of the \icond to soldering temperature}: The prepared inner conductor pieces are screwed into threaded holes (M4x0.5) in a copper block sitting on a hot plate heated to \SI{200}{\degreeCelsius} (as measured by a thermocouple attached to the copper block)
     \item \textbf{Ultrasonic soldering process}:
     The temperature of the ultrasonic soldering iron is set to \SI{200}{\degreeCelsius} and left to heat up. 
     The ultrasonic soldering iron uses special solder (CERASOLZER ALU-200, \SI{1.6}{\milli\meter} diameter) that melts at this temperature. The tip of the soldering iron is placed at the surface of the \icond piece  where the two wire ends bend around the surface.  Only a small quantity of solder is needed at the joint. 
      Apply ultrasonic power by pressing down on the step-on switch below the table.  The solder will start to wet the surfaces after applying the ultrasonic power. 
   \end{enumerate}
\subsubsection{Assembly database}
\label{subsubsec:provenance}
We maintained a database to record the assembly history and quality checks for each probe. 
It is accessed by the probe's unique identification number. 
It includes the person responsible for assembling the probe and the date the assembly was completed.
After assembly, the series LC circuit in each probe was tuned to 61.74 MHz by observing the reflection plot when connected to a vector network analyzer. 
Tuning occurs by varying the probe capacitance with a specially designed tool through the probe's end cap. 
The impedance is given by the attenuation of the reflected signal at resonance. 
Tuning this parameter requires opening the probe and deforming a few-turn coil parallel to the series resonant circuit. 
The database includes all plots and analyzed data for the probes. 
It is a reference for the initial probe qualities once the probes are installed at Fermilab.

%% file: main.bbl
\begin{thebibliography}{10}
\expandafter\ifx\csname url\endcsname\relax
  \def\url#1{\texttt{#1}}\fi
\expandafter\ifx\csname urlprefix\endcsname\relax\def\urlprefix{URL }\fi
\expandafter\ifx\csname href\endcsname\relax
  \def\href#1#2{#2} \def\path#1{#1}\fi

\bibitem{PhysRevD2003}
G.~W. Bennett, et~al.,
  \href{https://link.aps.org/doi/10.1103/PhysRevD.73.072003}{{Final report of
  the E821 muon anomalous magnetic moment measurement at BNL}}, Phys. Rev. D 73
  (2006) 072003.
\newblock \href {https://doi.org/10.1103/PhysRevD.73.072003}
  {\path{doi:10.1103/PhysRevD.73.072003}}.
\newline\urlprefix\url{https://link.aps.org/doi/10.1103/PhysRevD.73.072003}

\bibitem{PRIGL1996}
R.~Prigl, U.~Haeberlen, K.~Jungmann, G.~zu~Putlitz, P.~von Walter,
  \href{http://www.sciencedirect.com/science/article/pii/0168900296374937}{{"A
  high precision magnetometer based on pulsed NMR"}}, Nuclear Instruments and
  Methods in Physics Research Section A: Accelerators, Spectrometers, Detectors
  and Associated Equipment 374~(1) (1996) 118 -- 126.
\newblock \href {https://doi.org/https://doi.org/10.1016/0168-9002(96)37493-7}
  {\path{doi:https://doi.org/10.1016/0168-9002(96)37493-7}}.
\newline\urlprefix\url{http://www.sciencedirect.com/science/article/pii/0168900296374937}

\bibitem{Phillips_1977}
W.~D. Phillips, W.~E. Cooke, D.~Kleppner,
  \href{https://dx.doi.org/10.1088/0026-1394/13/4/005}{Magnetic moment of the
  proton in h2o in bohr magnetons}, Metrologia 13~(4) (1977) 179.
\newblock \href {https://doi.org/10.1088/0026-1394/13/4/005}
  {\path{doi:10.1088/0026-1394/13/4/005}}.
\newline\urlprefix\url{https://dx.doi.org/10.1088/0026-1394/13/4/005}

\bibitem{TDR2015}
{Muon (g-2) Technical Design Report} (2018).
\newblock \href {http://arxiv.org/abs/1501.06858} {\path{arXiv:1501.06858}}.

\bibitem{HONG2021107020}
R.~Hong, et~al.,
  \href{https://www.sciencedirect.com/science/article/pii/S1090780721001099}{Systematic
  and statistical uncertainties of the hilbert-transform based high-precision
  fid frequency extraction method}, Journal of Magnetic Resonance 329 (2021)
  107020.
\newblock \href {https://doi.org/https://doi.org/10.1016/j.jmr.2021.107020}
  {\path{doi:https://doi.org/10.1016/j.jmr.2021.107020}}.
\newline\urlprefix\url{https://www.sciencedirect.com/science/article/pii/S1090780721001099}

\bibitem{Corrodi_2020}
S.~Corrodi, et~al.,
  \href{https://doi.org/10.1088/1748-0221/15/11/p11008}{{Design and performance
  of an in-vacuum, magnetic field mapping system for the Muon g-2 experiment}},
  Journal of Instrumentation 15~(11) (2020) P11008--P11008.
\newblock \href {https://doi.org/10.1088/1748-0221/15/11/p11008}
  {\path{doi:10.1088/1748-0221/15/11/p11008}}.
\newline\urlprefix\url{https://doi.org/10.1088/1748-0221/15/11/p11008}

\bibitem{Flay_2021}
D.~Flay, et~al.,
  \href{https://dx.doi.org/10.1088/1748-0221/16/12/P12041}{High-accuracy
  absolute magnetometry with application to the fermilab muon g-2 experiment},
  Journal of Instrumentation 16~(12) (2021) P12041.
\newblock \href {https://doi.org/10.1088/1748-0221/16/12/P12041}
  {\path{doi:10.1088/1748-0221/16/12/P12041}}.
\newline\urlprefix\url{https://dx.doi.org/10.1088/1748-0221/16/12/P12041}

\bibitem{farooq_absolute_2020}
M.~Farooq, et~al., Absolute {{Magnetometry}} with {$^{3}$}{{He}}, Physical
  Review Letters 124~(22) (2020) 223001.
\newblock \href {https://doi.org/10.1103/PhysRevLett.124.223001}
  {\path{doi:10.1103/PhysRevLett.124.223001}}.

\bibitem{CODATA2018}
E.~Tiesinga, P.~J. Mohr, D.~B. Newell, B.~N. Taylor,
  \href{https://link.aps.org/doi/10.1103/RevModPhys.93.025010}{{CODATA
  recommended values of the fundamental physical constants: 2018}}, Rev. Mod.
  Phys. 93 (2021) 025010.
\newblock \href {https://doi.org/10.1103/RevModPhys.93.025010}
  {\path{doi:10.1103/RevModPhys.93.025010}}.
\newline\urlprefix\url{https://link.aps.org/doi/10.1103/RevModPhys.93.025010}

\bibitem{Keeler}
J.~Keeler, {Understanding NMR Spectroscopy}, Wiley, 2002.

\bibitem{Bloch1946}
F.~Bloch, \href{https://link.aps.org/doi/10.1103/PhysRev.70.460}{{Nuclear
  Induction}}, Phys. Rev. 70 (1946) 460--474.
\newblock \href {https://doi.org/10.1103/PhysRev.70.460}
  {\path{doi:10.1103/PhysRev.70.460}}.
\newline\urlprefix\url{https://link.aps.org/doi/10.1103/PhysRev.70.460}

\bibitem{MBR}
{{MBR} electronics}, \href{http://www.sonicsolder.com/}{{Ultrasonic soldering}}
  (2025).
\newline\urlprefix\url{http://www.sonicsolder.com/}

\bibitem{Minicircuit:Catalog}
Mini-circuits product catalog,
  \url{https://www.minicircuits.com/pdfs/Complete%20Catalog_2021_Digital.pdf}
  (2021).

\bibitem{flowers_correcting_1995}
J.~Flowers, P.~Franks, B.~Petley, Correcting high precision pulsed {{NMR}} flux
  density measurements for asymmetries in the lineshape, IEEE Trans. Instrum.
  Meas. 44~(2) (1995) 488--490.
\newblock \href {https://doi.org/10.1109/19.377888}
  {\path{doi:10.1109/19.377888}}.

\bibitem{gemmel_ultra-sensitive_2010}
C.~Gemmel, et~al., Ultra-sensitive magnetometry based on free precession of
  nuclear spins, Eur. Phys. J. D 57~(3) (2010) 303--320.
\newblock \href {https://doi.org/10.1140/epjd/e2010-00044-5}
  {\path{doi:10.1140/epjd/e2010-00044-5}}.

\bibitem{PhysRevA.103.042208}
T.~Albahri, et~al.,
  \href{https://link.aps.org/doi/10.1103/PhysRevA.103.042208}{Magnetic-field
  measurement and analysis for the muon $g\ensuremath{-}2$ experiment at
  fermilab}, Phys. Rev. A 103 (2021) 042208.
\newblock \href {https://doi.org/10.1103/PhysRevA.103.042208}
  {\path{doi:10.1103/PhysRevA.103.042208}}.
\newline\urlprefix\url{https://link.aps.org/doi/10.1103/PhysRevA.103.042208}

\bibitem{PhysRevD.103.072002}
T.~Albahri, et~al.,
  \href{https://link.aps.org/doi/10.1103/PhysRevD.103.072002}{Measurement of
  the anomalous precession frequency of the muon in the fermilab muon
  $g\ensuremath{-}2$ experiment}, Phys. Rev. D 103 (2021) 072002.
\newblock \href {https://doi.org/10.1103/PhysRevD.103.072002}
  {\path{doi:10.1103/PhysRevD.103.072002}}.
\newline\urlprefix\url{https://link.aps.org/doi/10.1103/PhysRevD.103.072002}

\bibitem{aguillard_detailed_2024}
D.~P. Aguillard, et~al., Detailed {{Report}} on the {{Measurement}} of the
  {{Positive Muon Anomalous Magnetic Moment}} to 0.20 ppm (Feb. 2024).
\newblock \href {http://arxiv.org/abs/2402.15410} {\path{arXiv:2402.15410}},
  \href {https://doi.org/10.48550/arXiv.2402.15410}
  {\path{doi:10.48550/arXiv.2402.15410}}.

\bibitem{ops_paper}
{Muon g-2 Collaboration}, P.~D. Aguillared, et~al., {Operations of the Muon g-2
  experiment at Fermilab (E989)}{ \it in preperation}.

\bibitem{muon_g2_collaboration_beam_2021}
{Muon g-2 Collaboration}, T.~Albahri, et~al., Beam dynamics corrections to the
  {{Run-1}} measurement of the muon anomalous magnetic moment at {{Fermilab}},
  Phys. Rev. Accel. Beams 24~(4) (2021) 044002.
\newblock \href {https://doi.org/10.1103/PhysRevAccelBeams.24.044002}
  {\path{doi:10.1103/PhysRevAccelBeams.24.044002}}.

\end{thebibliography}
